\documentclass[sigconf]{acmart}
\renewcommand\footnotetextcopyrightpermission[1]{} % removes footnote with conference information in first column
\usepackage{booktabs} % For formal tables
\usepackage{tabularx}
\usepackage{xspace}
\usepackage{enumitem}
\usepackage{xcolor,colortbl}
\usepackage{amssymb}
\usepackage{amsmath}
\usepackage{multirow}
\usepackage{hyperref}
\usepackage{graphicx}
\usepackage[center]{subfigure}
\usepackage[ruled,linesnumbered]{algorithm2e} % algorithm
\usepackage[]{algorithmic} % algorithm
\usepackage{tablefootnote}
\usepackage{threeparttable}
\usepackage{makecell}
\usepackage[font={small}]{caption}

\setcopyright{rightsretained}

\newcolumntype{L}[1]{>{\raggedright\let\newline\\\arraybackslash\hspace{0pt}}m{#1}}
\newcolumntype{C}[1]{>{\centering\let\newline\\\arraybackslash\hspace{0pt}}m{#1}}
\newcolumntype{R}[1]{>{\raggedleft\let\newline\\\arraybackslash\hspace{0pt}}m{#1}}

\acmDOI{10.475/1234}

\acmISBN{123-4567-24-567/08/06}

\acmConference[WWW'18]{The Web Conference}{April 2018}{Lyon, France}
\acmYear{2018}
\copyrightyear{2018}

\acmSubmissionID{123-A12-B3}

\begin{document}
\title{Fast and Accurate Random Walk with Restart on \\Dynamic Graphs with Guarantees}

%\author{Anonymous}
%\affiliation{
%\institution{Anonymous}
%}
%\email{Anonymous@Anonymous}

\author{Minji Yoon}
\affiliation{
	\institution{Seoul National University}
}
\email{riin55@snu.ac.kr}

\author{WooJeong Jin}
\affiliation{
	\institution{Seoul National University}
}
\email{woojung211@snu.ac.kr}

\author{U Kang}
\affiliation{
	\institution{Seoul National University}
}
\email{ukang@snu.ac.kr}

% The default list of authors is too long for headers
%\renewcommand{\shortauthors}{Anonymous}

% make the title area

\newtheorem{observation}{Observation}
\newtheorem{problem}{Problem}
\newtheorem{algo}{Algorithm}
\newtheorem{property}{Property}

\newcommand{\methodD}{\textsc{OSP}\xspace}
\newcommand{\methodDA}{\textsc{OSP-T}\xspace}
\newcommand{\methodC}{\textsc{CPI}\xspace}

\newcommand{\methodCI}{\textsc{CPI}\xspace} %exact

\newcommand{\mat}[1]{\mathbf{#1}}
\newcommand{\set}[1]{\mathbf{#1}}
\newcommand{\vect}[1]{\mathbf{#1}}

\renewcommand{\r}{\vect{r}} % for an rwr vector
\newcommand{\rst}{\r_{\text{stranger}}} %
\newcommand{\trst}{\vect{\tilde{r}}_{\text{stranger}}} %
\newcommand{\rnb}{\r_{\text{neighbor}}} %
\newcommand{\rtilde}{\vect{\tilde{r}}}
\newcommand{\trnb}{\vect{\tilde{r}}_{\text{neighbor}}} %
\newcommand{\rfm}{\r_{\text{family}}} %

\newcommand{\p}{\vect{p}} % for the pagerank vector
\newcommand{\q}{\vect{q}} % for the query vector
\newcommand{\x}{\vect{x}} % for the query vector
\newcommand{\xtilde}{\vect{\tilde{x}}} % for the query vector

\newcommand{\A}{\mat{A}} % normalized adjacency matrix
\newcommand{\NA}{\mat{\tilde{A}}} % normalized adjacency matrix
\newcommand{\NAT}{\mat{\tilde{A}}^{\top}} % normalized
\newcommand{\B}{\mat{B}} % normalized adjacency matrix
\newcommand{\NB}{\mat{\tilde{B}}} % normalized adjacency matrix
\newcommand{\NBT}{\mat{\tilde{B}}^{\top}} % normalized

\newcommand{\BA}{\mat{\bar{A}}}
\newcommand{\BB}{\mat{\bar{B}}}
\newcommand{\BAT}{\mat{\bar{A}}^{\top}}
\newcommand{\BBT}{\mat{\bar{B}}^{\top}}
\newcommand{\bq}{\vect{\bar{q}}}

\newcommand{\DA}{\Delta\mat{A}} % normalized adjacency matrix
\newcommand{\DAT}{{(\Delta\mat{A})}^{\top}} % normalized
\newcommand{\DNA}{\Delta\mat{\tilde{A}}} % normalized
\newcommand{\DNAT}{{\Delta\mat{\tilde{A}}}^{\top}} % normalized

\newcommand{\rev}[1]{\textcolor{blue}{#1}}
\newcommand{\red}[1]{\textcolor{red}{#1}} 

\begin{abstract}
Given a time-evolving graph, how can we track similarity between nodes in a fast and accurate way, with theoretical guarantees on the convergence and the error?
Random Walk with Restart (RWR) is a popular measure to estimate the similarity between nodes and has been exploited in numerous applications. %including ranking, anomaly detection, and community detection.
%There have been much efforts to calculate RWR efficiently in static graphs.
%However, RWR on dynamic graphs has been relatively less studied comparing to RWR on static graphs.
%Considering that many real-world graphs are dynamic with frequent insertion/deletion of edges, tracking RWR scores on dynamic graphs in an efficient way is an important issue.
Many real-world graphs are dynamic with frequent insertion/deletion of edges; thus, tracking RWR scores on dynamic graphs in an efficient way has aroused much interest among data mining researchers.
Recently, dynamic RWR models based on the propagation of scores across a given graph have been proposed, and have succeeded in outperforming previous other approaches to compute RWR dynamically.
However, those models fail to guarantee exactness and convergence time for updating RWR in a generalized form.

In this paper, we propose \methodD, a fast and accurate algorithm for computing dynamic RWR with insertion/deletion of nodes/edges in a directed/undirected graph.
When the graph is updated, \methodD first calculates %the difference in RWR scores around the modified edges,
offset scores around the modified edges, propagates the offset scores across the updated graph, % with fast convergence,
and then merges them with the current RWR scores to get updated RWR scores. %to counteract the difference.
We prove the exactness of \methodD and introduce \methodDA, a version of \methodD which regulates a trade-off between accuracy and computation time by using error tolerance $\epsilon$.
Given restart probability $c$, \methodDA guarantees to return RWR scores with $O(\epsilon/c)$ error in $O(\log_{(1-c)}(\frac{\epsilon}{2}))$ iterations.
Through extensive experiments, we show that
\methodD tracks RWR exactly up to $4605\times$ faster than existing static RWR method on dynamic graphs, and
\methodDA requires up to $15\times$ less time with $730\times$ lower $L1$ norm error and $3.3\times$ lower rank error than other state-of-the-art dynamic RWR methods.
\end{abstract}

%
% The code below should be generated by the tool at
% http://dl.acm.org/ccs.cfm
% Please copy and paste the code instead of the example below.
%

%\ccsdesc[300]{Information systems~Data mining}


\maketitle

\section{Introduction}
\label{sec:introduction}
\begin{table*}[]
	\small
	\vspace{-3mm}
	\centering
	\caption{
		Comparison of our proposed \methodD, \methodDA and existing methods for RWR computation on dynamic graphs.
		\methodD computes dynamic RWR exactly with reasonable time, while \methodDA shows the best trade-off between speed and accuracy among approximate methods.
		\methodD and \methodDA apply to the most general settings with guarantees.
	}
	\label{salesman}
	\vspace{1mm}
	\begin{tabular}{llllll}
		\hline
		\textbf{Method} & \textbf{Speed} & \textbf{Accuracy} & \textbf{Coverage} & \textbf{Accuracy Bound}& \textbf{Time complexity model} \\ \hline
		%\methodC~\cite{YoonJK17}&  Slow&  High&  Directed/Undirected graph & Yes & General \\
		TrackingPPR~\cite{ohsaka2015efficient}&  Fast & Low& Undirected graph& No &  Only with insertion of edges\\
		LazyForward~\cite{zhang2016approximate}& Fast &  Low&  Undirected graph& No & Only with undirected graph \\ \hline
		\textbf{\methodD}&  \textbf{Medium}&  \textbf{High}& \textbf{Directed/Undirected graph}& \textbf{Yes} &  \textbf{General}\\
		\textbf{\methodDA}&  \textbf{Faster}&  \textbf{Medium}& \textbf{Directed/Undirected graph}& \textbf{Yes} & \textbf{General} \\ \hline
	\end{tabular}
	\vspace{1mm}
\end{table*}

Identifying similarity score between two nodes in a graph has been recognized as a fundamental tool to analyze the graph~\cite{zhu2013local, whang2013overlapping, tong2006center, sun2005neighborhood}
and has been exploited in various graph mining tasks~\cite{fujiwara2012fast, chakrabarti2011index, antonellis2007query}. % to gain insights about the given graph
Among numerous methods~\cite{jeh2002simrank, pan2004automatic, lin2009matchsim} to measure the similarity, random walk with restart (RWR)~\cite{pan2004automatic} has aroused considerable attention due to its ability to account for the global network structure from a particular user's point of view. %multi-faceted relationship between nodes in a graph~\cite{tong2006center}
RWR has been widely used in various applications across different domains including ranking, link prediction~\cite{JINJK17}, and recommendation~\cite{DBLP:journals/corr/abs-1708-09088}.
To avoid expensive costs incurred by RWR computation, various methods have been proposed to calculate RWR scores efficiently, and the majority of them have focused on static graphs~\cite{shin2015bear, JungPSK17, wang2016hubppr,DBLP:conf/icdm/JungJSK16,DBLP:journals/tods/JungSSK16}.
However, many real-world graphs are dynamic.
For example, in an online social network of Facebook which has more than $1.3$ billion users, $5$ new users are added every second,
and the total number of websites on the world wide web fluctuates up to $600$ thousands around $60$ trillion web pages every second~\cite{ohsaka2015efficient}. %~\footnote{http://www.internetlivestats.com/}
Thus it is an indispensable task to track RWR scores on time-evolving real-world graphs.

Various approaches have been proposed to handle dynamic RWR problem efficiently.
Chien et al.~\cite{chien2004link} introduced an approximate aggregation/disaggregation method which updates RWR scores only around modified edges.
Bahmani et al.~\cite{bahmani2010fast} applied the Monte-Carlo method~\cite{jeh2003scaling} on the dynamic RWR problem. %with running time analysis in a random edge arrival order model.
Recently, score propagation models were proposed by Ohsaka et al.~\cite{ohsaka2015efficient} and Zhang et al.~\cite{zhang2016approximate};
Ohsaka et al. proposed TrackingPPR which propagates scores using Gauss-Southwell algorithm;
Zhang et al. proposed LazyForward which optimizes the initial step from TrackingPPR and propagates scores using Forward Push algorithm.
%in these models, RWR score difference made by a graph modification is calculated at first and propagated across the graph.
They succeed in outperforming the previous approaches in both running time and accuracy~\cite{ohsaka2015efficient, zhang2016approximate}.
However, they fail to provide theoretical analysis of accuracy bound.
%theoretical analysis of time complexity and accuracy bound in a generalized form.
Furthermore, they narrow down the scope of their analyses on time complexity to graph modifications only with insertion of edges or graph modifications on undirected graphs.

\begin{figure}[!t]
	\centering
	\vspace{1mm}
	\includegraphics[width=1\linewidth]{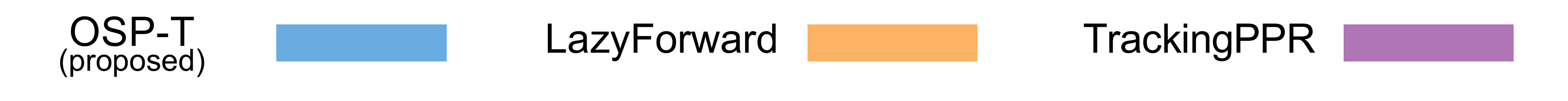}\\
	\vspace{-1.5mm}
	\includegraphics[width=.7\linewidth]{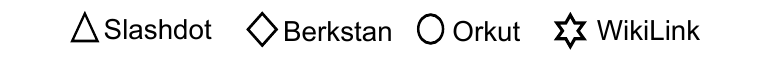}\\
	\vspace{-1.5mm}
	\hspace{-5mm}
	\subfigure[Accuracy on L1 norm of error]
	{
		\includegraphics[width=.5\linewidth]{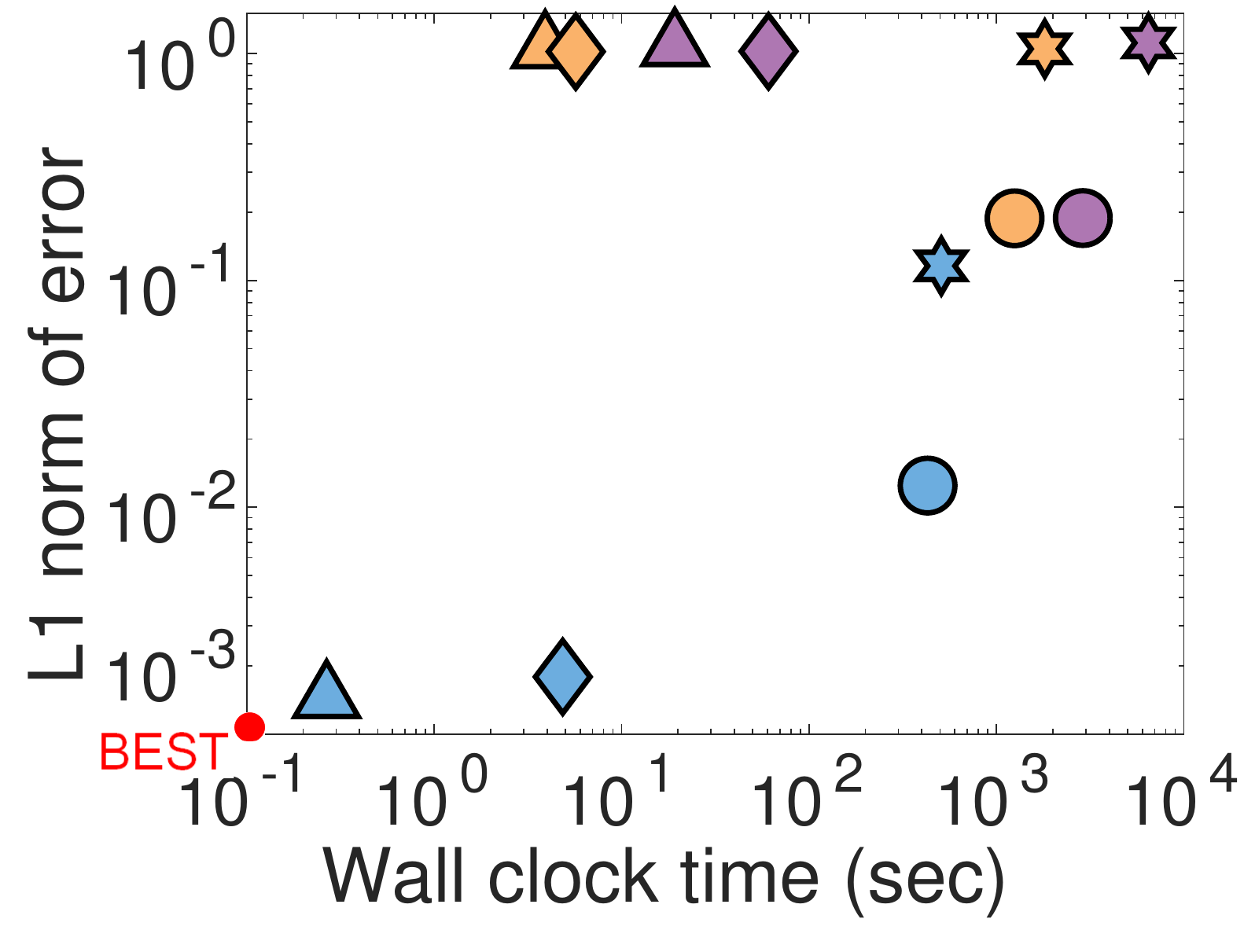}
	}
	\hspace{-2mm}
	\subfigure[Accuracy on Rank]
	{
		\includegraphics[width=.49\linewidth]{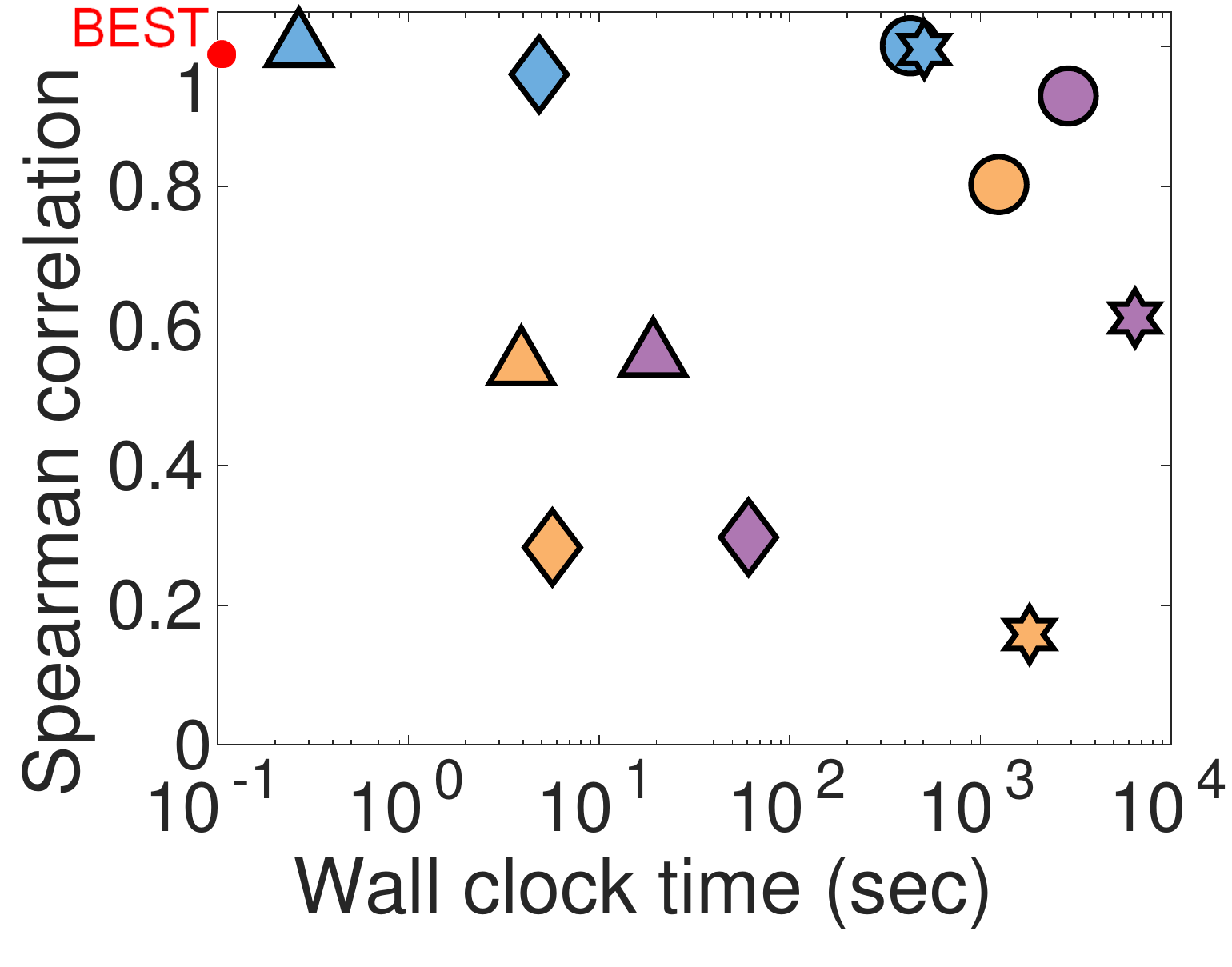}
	}
	\caption
	{
		Trade-off between accuracy and time:
		\methodDA shows the best trade-off between speed and accuracy among approximate methods for dynamic RWR.
	}
	\label{fig:scatter}
\end{figure}

In this paper, we propose \methodD~(Offset Score Propagation for RWR), a fast and accurate method model for computing RWR scores on dynamic graphs.
\methodD is based on cumulative power iteration (\methodC)~\cite{YoonJK17} which interprets an RWR problem as propagation of scores from a seed node across a graph in an iterative matrix-vector multiplication form.
When the graph is updated, \methodD first calculates offset scores made around the updated edges, and then propagates the offset scores across the modified graph using \methodC.
The small size of the offset scores leads to fast convergence.
Then \methodD merges the result of the propagation with the current RWR scores to get an updated RWR scores.
Unlike the previous propagation models~\cite{ohsaka2015efficient, zhang2016approximate}, \methodD gives exact updated RWR scores. %in a generalized form.
We also propose \methodDA, a version of \methodD, with provable error bound and running time:
given restart probability $c$ and error tolerance $\epsilon$, \methodDA computes RWR scores with $O(\epsilon/c)$ error in $O(\log_{(1-c)}(\frac{\epsilon}{2}))$ iterations in dynamic graphs.
Through extensive experiments with various real-world graphs, we demonstrate the superior performance of \methodD and \methodDA over existing methods.
Table~\ref{salesman} and Figure~\ref{fig:scatter} show a comparison of \methodD, \methodDA, and existing methods.
The main contributions of this paper are the followings:

\begin{itemize}
	\item{
		\textbf{Algorithm.}
		We introduce \methodD, a fast and accurate method to compute RWR on dynamic graphs.
		We also propose \methodDA, a version of \methodD which regulates a trade-off between accuracy and computation time by using an error tolerance parameter.
	}
	\item{
		\textbf{Analysis.}
		We present a theoretical analysis on exactness of \methodD, and time complexity and error bound of \methodDA.
		Our analysis is applicable to general dynamic graphs:
		insertion / deletion of nodes / edges in directed / undirected graphs.
	}
	\item{
		\textbf{Experiment.}
		We present extensive empirical evidences for the performance of \methodD and \methodDA using various dynamic real-world graphs.
		%We compare \methodDA with the state-of-the-art RWR methods for dynamic graphs.
		We show that \methodD tracks RWR exactly up to $4605\times$ faster than existing static RWR method, and
		\methodDA requires up to $15\times$ less time with $730\times$ lower $L1$ norm error and $3.3\times$ lower rank error than other dynamic RWR methods.
	}
\end{itemize}

The rest of the paper is organized as follows.
In Section~\ref{sec:preliminaries}, we describe preliminaries on RWR and \methodC.
In Section~\ref{sec:proposed_method}, we present the proposed model \methodD and \methodDA in detail along with theoretical analyses.
After presenting our experimental results in Section~\ref{sec:experiments}, we provide a review on related works in Section~\ref{sec:related_works} and conclude in Section~\ref{sec:conclusion}.
The symbols frequently used in this paper are summarized in Table~\ref{tab:symbols}.

\section{Preliminaries}
\label{sec:preliminaries}
In this section, we briefly review RWR~\cite{tong2006fast}, and explain \methodC~\cite{YoonJK17}, an iterative method for RWR computation.
\vspace{-3mm}
\subsection{Random Walk with Restart}\label{sec:rwr}
Random walk with restart (RWR)~\cite{tong2006fast} estimates each node's relevance with regard to a given seed node $s$ in a graph.
RWR assumes a random surfer who starts at node $s$.
In each step, the surfer follows edges with probability $1-c$, or jumps to the seed node with probability $c$.
The surfer chooses an edge to move on with uniform probability among all current outgoing edges.
%Then the expected frequency of visiting node $i$ by the surfer, denoted by $\r(i)$, becomes the RWR score of node $i$.
The vector $\r$ representing each node's visiting probability satisfies the following equation:
%\vspace{-7mm}
\begin{equation}
\small
\r=(1-c) \NAT \r + c \q
\end{equation}
where $\NA$ is a row-normalized adjacency matrix and $\q$ is a starting vector whose $s$th element is $1$ and other elements are $0$.
%whose $(i,j)$th entry is the probability that a surfer at node $i$ chooses the edge to node $j$ among all of its outgoing edges.
%$c$ is the restart probability and $\q$ is a starting vector whose $s$th element is $1$ and other elements are $0$.

\begin{table}[!t]
	\centering
	\small
	\caption{Table of symbols.}
	\begin{tabular}{cl}
		\toprule
		\textbf{Symbol} & \textbf{Definition} \\
		\midrule
		$G$ & {\small input graph}\\
		$\Delta G$ & {\small update in graph}\\
		$n,m$ & {\small numbers of nodes and edges in $G$}\\
		%$m$ & {\small number of edges in $G$}\\
		$s$ & {\small seed node (= query node, source node)}\\
		$c$ & {\small restart probability}\\
		$\epsilon$ & {\small error tolerance} \\
		$\q$ & {\small ($n\times1$) starting vector whose $s$th element is $1$} \\&{and other elements are $0$}\\
		%$\A$ & {\small ($n \times n$) adjacency matrix of $G$}\\
		$\NA$ & {\small ($n \times n$) row-normalized adjacency matrix of $G$}\\
		%$\B$ & {\small ($n \times n$) adjacency matrix of $G+\Delta G$}\\
		$\NB$ & {\small ($n \times n$) row-normalized adjacency matrix of $G+\Delta G$}\\
		$\DA$ & {\small ($n \times n$) difference between $\NA$ and $\NB$ ($= \NB - \NA$)}\\
		$\r_{\text{old}}$ & {\small ($n\times1$) RWR vector on $G$}\\
		$\r_{\text{new}}$ &{\small ($n\times1$) updated RWR vector on $G+\Delta G$} \\
		$\q_{\text{offset}}$ & {\small ($n\times1$) offset seed vector}\\
		$\x_{\text{offset}}^{(i)}$ & {\small ($n\times1$) interim offset score vector at $i$th iteration in \methodD}\\
		$\r_{\text{offset}}$ & {\small ($n\times1$) offset score vector}\\
		\bottomrule
	\end{tabular}
	\label{tab:symbols}
	\vspace{1mm}
\end{table}

\subsection{\methodC: Cumulative Power Iteration}\label{sec:cpi}
Cumulative power iteration (\methodC)~\cite{YoonJK17} interprets an RWR problem as a propagation of scores across a graph in an iterative matrix-vector multiplication form:
a score $c$ is generated from a seed node at first; at each iteration, scores are divided and propagated evenly into out-edges with decaying coefficient $1-c$.
$\x^{(i)}$ is an interim score vector computed from the iteration $i$ and has scores propagated across nodes at $i$th iteration as entries.
When multiplied with $(1-c)\NAT$, scores in $\x^{(i)}$ are propagated into their outgoing neighbors, and the propagated scores are stored in $\x^{(i+1)}$.
\methodC accumulates interim score vectors $\x^{(i)}$ to get the final RWR score vector $\r_{\text{\methodC}}$ as follows.
\small
\begin{align*}
\x^{(0)} &= \q_{\text{\methodC}} \\
\x^{(i)} &= (1-c)\NA^{\top}\x^{(i-1)} = \left((1-c)\NAT\right)^{i}\q_{\text{\methodC}} \\
\r_{\text{\methodC}} &= \sum_{i=0}^{\infty}\x^{(i)} = \sum_{i=0}^{\infty}\left((1-c)\NAT\right)^{i}\q_{\text{\methodC}}
\end{align*}
\normalsize

\noindent$\q_{\text{\methodC}}$ is a seed vector which contains initial scores for propagation.
For RWR computation, $\q_{\text{\methodC}} = c\q$ is set with an initial score $c$ at seed index $s$.
Unlike other propagation methods such as Gauss-Southwell algorithm~\cite{ohsaka2015efficient} and Forward Local Push algorithm~\cite{zhang2016approximate},
\methodC computes RWR with accuracy assurance and general time complexity model.
Thus we propose dynamic RWR method \methodD and \methodDA using \methodC to provide 
theoretical guarantees on the error and the convergence.

\section{Proposed Method}
\label{sec:proposed_method}
In this section, we describe our proposed method \methodD for tracking RWR on dynamic graphs, and introduce \methodDA, a version of \methodD which regulates a trade-off between accuracy and computation time on dynamic graphs.

\vspace{-1mm}
\subsection{\methodD: Offset Score Propagation}
\label{subsec:dynamic}

In \methodC, scores are propagated following underlying edges, and the score accumulated in each node becomes RWR score of the node in a given graph.
In other words, RWR scores of nodes are determined by distribution of edges.
In this sense, when the graph $G$ is updated with an insertion/deletion of edges ($\Delta G$), propagation of scores around $\Delta G$ is changed:
with insertion of an edge $e = (u,v)$, score $x_{u}$ from the node $u$ would be propagated into each out-edges with smaller scores $\frac{x_{u}}{d_{u}+1}$ where $d_{u}$ is the number of out-edges of node $u$ before $\Delta G$;
with deletion of an edge $e = (u,v)$, score $x_u$ from the node $u$ would be propagated into each out-edges with higher scores $\frac{x_u}{d_u-1}$.
Then, these changes are propagated, affect the previous propagation pattern across the whole graph, and finally lead to a different RWR vector $r_{\text{new}}$ of the updated graph $G+\Delta G$ from $r_{\text{old}}$ of the original graph $G$.
Based on this observation, \methodD first calculates an offset seed vector $\q_{\text{offset}}$ and propagates the offset scores across the updated graph $G+\Delta G$ using \methodC to get an offset score vector $\r_{\text{offset}}$.
Finally, \methodD adds up $\r_{\text{old}}$ and $\r_{\text{offset}}$ to get the final RWR score vector $\r_{\text{new}}$ as follows:
\small
\begin{align*}
	&\q_{\text{offset}} \leftarrow (1-c)(\NBT - \NAT)\r_{\text{old}} = (1-c)\DAT\r_{\text{old}} \\
	&\x_{\text{offset}}^{(i)} \leftarrow ((1-c)\NBT)^{i}\q_{\text{offset}}\\
	&\r_{\text{offset}} \leftarrow \sum_{i=0}^{\infty}\x_{\text{offset}}^{(i)} = \sum_{i=0}^{\infty}((1-c)\NBT)^{i}\q_{\text{offset}} \\
	&\r_{\text{new}} \leftarrow \r_{\text{old}} + \r_{\text{offset}}
\end{align*}
\normalsize
\noindent where $\NA$ is a row-normalized adjacency matrix of $G$,
$\NB$ is a row-normalized adjacency matrix of $G+\Delta G$, and
$\DA =\NB-\NA$ is the difference between $\NB$ and $\NA$.
Before proving the exactness of $\r_{\text{new}}$ computed by \methodD, we first show the convergence of $\r_{\text{offset}}$ in Lemma~\ref{lemma:convergence}.
\begin{lemma}[Convergence of $\r_{\text{offset}}$]
	\label{lemma:convergence}
	$\r_{\text{offset}}$ converges to a constant value.
	\begin{proof}
		%Let $\x^{(i)}$ denote an interim score vector at $i$th iteration when
		$\r_{\text{offset}}$ is represented in \methodD as follows:
		\small
		\begin{align*}
		%\x^{(i)} &= ((1-c)\NBT)^{i} \q_{\text{offset}}\\
		\r_{\text{offset}} &= \sum_{i=0}^{\infty}\x_{\text{offset}}^{(i)} = \sum_{i=0}^{\infty}((1-c)\NBT)^{i}\q_{\text{offset}} \\
		&= \sum_{i=0}^{\infty}(1-c)^{i}(\NBT)^{i}(1-c)(\NBT - \NAT)\r_{\text{old}}\\
		&= \sum_{i=0}^{\infty}((1-c)\NBT)^{(i+1)}\r_{\text{old}} - \sum_{i=0}^{\infty}(1-c)^{(i+1)}(\NBT)^{i}\NAT\r_{\text{old}}\\
		\end{align*}
		\normalsize
		Note that $\lVert\r_{\text{old}}\rVert_{1} = \lVert\NAT\rVert_{1} = \lVert\NBT\rVert_{1} = 1$ since $\r_{\text{old}}$ is an RWR score vector, and $\NA$ and $\NB$ are row-normalized stochastic matrices.
		Then $\sum_{i=0}^{\infty}((1-c)\NBT)^{(i+1)}r_{\text{old}}$ and $\sum_{i=0}^{\infty}(1-c)^{(i+1)}(\NBT)^{i}\NAT\r_{\text{old}}$ converge, and thus, $\r_{\text{offset}}$ also converges to a constant value.	
	\end{proof}
\end{lemma}
%\vspace{-3mm}
\noindent In Theorem~\ref{theorem:exactness}, we show that the sum of $\r_{\text{old}}$ and $\r_{\text{offset}}$ becomes the exact RWR score vector of the updated graph $G+\Delta G$.
Our result is the first exactness guarantee for propagation approaches~\cite{ohsaka2015efficient, zhang2016approximate} on dynamic graphs.
%\methodD guarantees its exactness while the previous propagating approaches~\cite{ohsaka2015efficient, zhang2016approximate} do not provide any accuracy analysis.

\begin{theorem}[Exactness of $\methodD$]
	\label{theorem:exactness}
	$\r_{\text{new}}$ computed by \methodD is the exact RWR score vector of the updated graph $G + \Delta G$.
	\begin{proof}
		For brevity, let $\BA \leftarrow (1-c)\NA, \BB \leftarrow (1-c)\NB, \bq \leftarrow c\q$ during this proof.
		Thus, the spectral radii of $\BA$ and $\BB$ become less than $1$ during this proof.
		\small
		\begin{align*}
		\!\!\!\r_{\text{offset}} &= \sum_{i=0}^{\infty}(\BBT)^i\q_{\text{offset}} \\
		&= \sum_{i=0}^{\infty}(\BBT)^i(\BBT -\BAT)\r_{\text{old}} \\
		&= \sum_{i=0}^{\infty}\left((\BBT)^i(\BBT -\BAT)\sum_{k=0}^{\infty}(\BAT)^{k}\bq\right) \\
		&= \sum_{i=0}^{\infty}\left((\BBT)^{i+1}\sum_{k=0}^{\infty}(\BAT)^{k}\bq - (\BBT)^i\sum_{k=0}^{\infty}(\BAT)^{k+1}\bq\right)
		\end{align*}
		\normalsize
		Note that $\r_{\text{old}}$ is represented as $\sum_{k=0}^{\infty}(\BAT)^{k}\bq$ in \methodC.
		
		The third summation $\sum_{k=0}^{\infty}(\BAT)^{k+1}\bq$  in the last equation is expressed as follows:
		\small
		\begin{align*}
			\sum_{k=0}^{\infty}(\BAT)^{k+1}\bq = ( \sum_{k=0}^{\infty}(\BAT)^{k}\bq ) - \bq
		\end{align*}
		\normalsize
		Using this equation, $\r_{\text{offset}}$ is
		\small
		\begin{align*}
		\!\!\!\r_{\text{offset}} &= \sum_{i=0}^{\infty}\left((\BBT)^{i+1}\sum_{k=0}^{\infty}(\BAT)^{k}\bq - (\BBT)^i\sum_{k=0}^{\infty}(\BAT)^{k}\bq + (\BBT)^{i}\bq\right) \\
		&= \sum_{i=0}^{\infty}(\BBT)^{i+1}\sum_{k=0}^{\infty}(\BAT)^k\bq - \sum_{i=0}^{\infty}(\BBT)^{i}\sum_{k=0}^{\infty}(\BAT)^k\bq + \sum_{i=0}^{\infty}(\BBT)^{i}\bq
		\end{align*}
		\normalsize
		The first term of the last equation, $\sum_{i=0}^{\infty}(\BBT)^{i+1}\sum_{k=0}^{\infty}(\BAT)^k\bq$  is expressed as follows:
		\small
		\begin{align*}
		\sum_{i=0}^{\infty}(\BBT)^{i+1}\sum_{k=0}^{\infty}(\BAT)^k\bq = \sum_{i=0}^{\infty}(\BBT)^{i}\sum_{k=0}^{\infty}(\BAT)^k\bq - \sum_{k=0}^{\infty}(\BAT)^{k}\bq
		\end{align*}
		\normalsize
		Then $\r_{\text{offset}}$ is expressed as follows:
		\small
		\begin{align*}
		\r_{\text{offset}} = -\sum_{k=0}^{\infty}(\BAT)^{k}\bq + \sum_{i=0}^{\infty}(\BBT)^{i}\bq
		\end{align*}
		\normalsize
		Note $\r_{\text{old}} = \sum_{k=0}^{\infty}(\BAT)^{k}\bq$ in \methodC. Then $\r_{\text{new}}$ becomes as follows:
		\small
		\begin{align*}
		\r_{\text{new}} &= \r_{\text{old}} + \r_{\text{offset}} \\
		&= \sum_{k=0}^{\infty}(\BAT)^{k}\bq -\sum_{k=0}^{\infty}(\BAT)^{k}\bq + \sum_{i=0}^{\infty}(\BBT)^{i}\bq \\
		&= \sum_{i=0}^{\infty}(\BBT)^{i}\bq
		\end{align*}
		\normalsize
		Note that RWR score vector of the updated graph $G+\Delta G$ is expressed as $\sum_{i=0}^{\infty}(\BBT)^{i}\bq$ in \methodC.
	\end{proof}
\end{theorem}

\noindent
Algorithm~\ref{alg:method:dynamic} describes how \methodD works.
\methodD first calculates a seed offset vector $\q_{\text{offset}}$ (line 1).
Then \methodD initializes RWR score vector $\r_{\text{offset}}$ and $\x_{\text{offset}}^{(0)}$ using the offset vector $\q_{\text{offset}}$ (line 2).
In $i$th iteration, scores in $\x_{\text{offset}}^{(i-1)}$ from the previous iteration ($i-1$) are propagated through $\NA+\DA$ with decaying coefficient $1-c$ (line 4).
Then, interim score vector $\x_{\text{offset}}^{(i)}$ is accumulated in $\r_{\text{offset}}$ (line 5).
\methodD stops when $\lVert \x_{\text{offset}}^{(i)} \rVert_{1} \le \epsilon$ which is a condition for the final score vector $\r_{\text{offset}}$ to converge (line 3).
%\methodCI stops iteration when the RWR score vector $\r$ is converged with an error tolerance $\epsilon$, and outputs $\r$ as a final RWR vector $\r_{\methodC}$.
%{\color{blue}Then \methodDA estimates RWR dynamically based on \methodCI.}
%Then \methodDA calls \methodCI to compute $\r_{\text{offset}}$ with an updated row-normalized adjacency matrix $\NA+\Delta\A$ and $\q_{\text{offset}}$ (line 2).
%We use the same error tolerance $\epsilon$ for \methodCI as that for \methodDA.
%\methodDA sets the convergence tolerance $\delta$ of \methodCI with its error tolerance $\epsilon$.
Finally, \methodD sums up $\r_{\text{old}}$ and $\r_{\text{offset}}$ (line 7).
To retrieve exact RWR scores, \methodD sets error tolerance $\epsilon$ to a very small value such as $10^{-9}$.
Using higher values for $\epsilon$, we propose an approximate method \methodDA which trades off the accuracy against the running time for users who put more priority on speed than accuracy in the following section.

\subsection{\methodDA: OSP with Tradeoff}
\label{subsec:dynamic-approx}

\methodDA is an approximate method for dynamic RWR computation which is based on \methodD.
As described in Algorithm~\ref{alg:method:dynamic}, the algorithm of \methodDA is the same as \methodD, but \methodDA regulates its accuracy and speed using error tolerance parameter $\epsilon$.
In the following, we analyze how much \methodDA sacrifices its accuracy and increases its speed when an error tolerance $\epsilon$ is given.
%Note that the exact method \methodD sets the error tolerance $\epsilon$ {\color{blue}close to $0$} in Algorithm~\ref{alg:method:dynamic}.
%In following, we provide theoretical analysis of \methodDA.
%We first evaluate the time complexity of \methodCI in Lemma~\ref{lemma:time_cpi} since \methodDA is based on \methodCI.
%, and then analyze \methodDA since \methodDA is based on \methodCI.

%\begin{algorithm} [t!]
%	\small
%	\begin{algorithmic}[1]
%		\caption{\methodCI Algorithm} \label{alg:method:cpi}
%		\REQUIRE row-normalized adjacency matrix: $\NA$, seed vector: $\q_{\text{\methodC}}$, \\restart probability: $c$, error tolerance: $\epsilon$
%		\ENSURE relevance score vector: $\r_{\text{\methodC}}$
%		\STATE set $\r = \vect{0}$ and $\x^{(0)} = \q_{\text{\methodC}}$
%		\FOR{iteration $i = 1$; $\lVert \x^{(i)} \rVert_{1} > \epsilon$; $i$++}
%		\STATE compute $\x^{(i)} \leftarrow (1-c)(\NAT\x^{(i-1)})$
%		\STATE compute $\r \leftarrow \r + \x^{(i)}$
%		\ENDFOR \label{alg:method:exact:iter:end}
%		\BlankLine
%		\RETURN $\r$
%	\end{algorithmic}
%\end{algorithm}

\begin{algorithm} [t!]
	\small
	\begin{algorithmic}[1]
		\caption{\methodD and \methodDA Algorithm}
		\label{alg:method:dynamic}
		\REQUIRE previous RWR score vector: $\r_{\text{old}}$, row-normalized adjacency matrix: $\NA$, update in $\NA$: $\DA$, restart probability: $c$, error tolerance: $\epsilon$
		\ENSURE updated RWR score vector: $\r_{\text{new}}$
		%\STATE $seed \leftarrow -1$ $\#$ denote PageRank
		\STATE set seed offset vector $\q_{\text{offset}} = (1-c)\DAT \r_{\text{old}}$
		%\STATE $\r_{\text{offset}} \leftarrow $ \methodCI($\NA + \DA$, $\q_{\text{offset}}$, $c$, $\epsilon$) $\#$ Algorithm~\ref{alg:method:cpi}
		\STATE set $\r_{\text{offset}} = \vect{0}$ and $\x_{\text{offset}}^{(0)} = \q_{\text{offset}}$
		\FOR{iteration $i = 1$; $\lVert \x_{\text{offset}}^{(i)} \rVert_{1} > \epsilon$; $i$++}
		\STATE compute $\x_{\text{offset}}^{(i)} \leftarrow (1-c)(\NA+\DA)^{\top}\x_{\text{offset}}^{(i-1)}$
		\STATE compute $\r_{\text{offset}} \leftarrow \r_{\text{offset}} + \x_{\text{offset}}^{(i)}$
		\ENDFOR \label{alg:method:exact:iter:end}
		\STATE $\r_{\text{new}} \leftarrow \r_{\text{old}} + \r_{\text{offset}}$
		\BlankLine
		\RETURN $\r_{\text{new}}$
	\end{algorithmic}
\end{algorithm}

%\vspace{-2mm}
%\begin{lemma}[Time Complexity of \methodCI; Lemma5 in ~\cite{YoonJK17}]
%	\label{lemma:time_cpi}
%	At each iteration, \methodCI takes $O(m)$ where $m$ is the number of edges in a given graph.
%	In total, \methodCI takes $O(m\log_{(1-c)}(\frac{\epsilon}{\lVert\q_{\text{\methodC}}\rVert_{1}}))$ time where $\q_{\text{\methodC}}$ is a seed vector and $\log_{(1-c)}(\frac{\epsilon}{\lVert\q_{\text{\methodC}}\rVert_{1}})$ indicates the number of iterations for convergence.
%	\vspace{-2mm}
%	\begin{proof}
%		For the completeness, we give a proof here.
%		\methodCI computes $\x^{(i)} = (1-c)\NAT\x^{(i-1)}$ for each iteration, and takes $O(m)$ time where $m$ is the number of nonzeros in $\NA$.
%		It also means the upper bound of number of edges visited in each iteration.
%		\methodCI stops the iteration with error tolerance $\epsilon$ when $\lVert \x^{(i)} \rVert_{1} = (1-c)^{i}\lVert\q_{\text{\methodC}}\rVert_{1} \le \epsilon$.
%		Then the number of iterations to be converged is $\log_{(1-c)}(\frac{\epsilon}{\lVert\q_{\text{\methodC}}\rVert_{1}})$ and total computation time becomes $O(m\log_{(1-c)}(\frac{\epsilon}{\lVert\q_{\text{\methodC}}\rVert_{1}}))$.
%	\end{proof}
%\end{lemma}

\vspace{-2mm}
\begin{theorem}[Time Complexity of \methodDA]
	\label{theorem:time_dynamic}
	With error tolerance $\epsilon$, \methodDA takes $O(m\log_{(1-c)}(\frac{\epsilon}{2}))$ where $m$ is the number of nonzeros in $\NA + \DA$.
	\vspace{-4mm}
	\begin{proof}
	$\NB$ denotes $\NA +\DA$, the row-normalized matrix for the updated graph.
	In each iteration, \methodDA computes $\x_{\text{offset}}^{(i)} = (1-c)\NBT\x_{\text{offset}}^{(i-1)}$, and takes $O(m)$ time where $m$ is the number of nonzeros in $\NBT$.
	It also means the upper bound of number of edges visited in each iteration.
	\methodDA stops the iteration with error tolerance $\epsilon$ when $\lVert \x_{\text{offset}}^{(i)} \rVert_{1} = \lVert((1-c)\NBT)^{i}\q_{\text{offset}}\rVert_{1}=(1-c)^{i}\lVert\q_{\text{offset}}\rVert_{1} \le \epsilon$.
	Note that $\NBT$ is a column stochastic matrix and $\lVert\NBT\rVert_{1}=1$.
	Then the number of iterations to be converged is $\log_{(1-c)}(\frac{\epsilon}{\lVert\q_{\text{offset}}\rVert_{1}})$ and total computation time becomes $O(m\log_{(1-c)}(\frac{\epsilon}{\lVert\q_{\text{offset}}\rVert_{1}}))$.
	The upper bound of $\lVert\q_{\text{offset}}\rVert_{1}$ is presented as follows:
	%\methodDA calls \methodCI with a seed vector $\q_{\text{\methodC}} = \q_{\text{offset}} = (1-c)(\NBT-\NAT)\r_{\text{old}}$.
	\small
	\begin{align*}
		\lVert\q_{\text{offset}}\rVert_{1} &= \lVert(1-c)(\NBT-\NAT)\r_{\text{old}}\rVert_{1}  \\
		&= (1-c)\lVert(\NBT-\NAT)\rVert_{1} \\
		&\leq (1-c)(\lVert\NBT\rVert_{1}+\lVert\NAT\rVert_{1}) = 2(1-c)
	\end{align*}
	\normalsize
	\vspace{-2mm}
	where $\lVert\r_{\text{old}}\rVert_{1} = \lVert\NAT\rVert_{1} = \lVert\NBT\rVert_{1} = 1$.
	{Then upper bounds of number of iterations and time are as follows:}
	%\methodDA calls \methodCI with $\q_{\text{offset}}$ for a seed vector $\q_{\text{\methodC}}$, and thus
	\small
	\begin{align*}
		O(iteration) = \log_{(1-c)}(\frac{\epsilon}{2(1-c)}) <  \log_{(1-c)}(\frac{\epsilon}{2})\\
		O(time) = m\log_{(1-c)}(\frac{\epsilon}{2(1-c)}) < m\log_{(1-c)}(\frac{\epsilon}{2})
	\end{align*}
	\normalsize
	%thus \methodDA iterates $O(\log_{(1-c)}(\frac{\epsilon}{2(1-c)}))$ times, and takes $O(m\log_{(1-c)}(\frac{\epsilon}{2})-m)$ time from Lemma~\ref{lemma:time_cpi}.
	Note that the upper bound of $\lVert\q_{\text{offset}}\rVert_{1}$ determines the upper bound of time complexity:
	$\lVert\q_{\text{offset}}\rVert_{1}$ is a denominator in $\log_{(1-c)}(\frac{\epsilon}{\lVert\q_{\text{offset}}\rVert_{1}})$ and the base of the logarithm is $1-c$ which is smaller than $1$.
	\end{proof}
\end{theorem}

\begin{table*}[t]
	\begin{threeparttable}[t]
	\centering
	\small
	\vspace{-1mm}
	\caption{
	Practical performance of \methodD and \methodDA on the LiveJournal dataset.
	Even though \methodC, \methodD and \methodDA share the same theoretical upper bound $O(m)$ of the number of visited edges per iteration,
	they show the different numbers of iterations and visited edges due to the different starting vectors $\q$ and error tolerance $\epsilon$.
	\methodD converges faster than \methodC with the help of smaller size of the starting vector, \methodDA $(\epsilon=5\times10^{-3})$ converges faster than \methodD $(\epsilon=10^{-9})$ with the help of higher error tolerance.
	Note that the total number of edges of LiveJournal is $34,681,189$.
	}
	\label{visited-edges}
	\begin{tabular}{C{11mm} | R{11mm} R{8mm} R{17mm} |R{13mm}R{8mm} R{17mm} |R{13mm} R{8mm}R{15mm}R{20mm}}\hline %{l|lll|lll|lll}
		\toprule
		\multirow{2}{*}{\shortstack{\#modified \\ edges}} & \multicolumn{3}{c|}{\textbf{\methodC}} & \multicolumn{3}{c|}{\textbf{\methodD}} & \multicolumn{3}{c}{\textbf{\methodDA}} \\
				   &   $\lVert\q_{\text{\methodC}}\rVert_{1} $ &  \textbf{\# iter}   &   \textbf{\#visited edges($\times10^3$)}     &   $\lVert\q_{\text{offset}} \rVert_{1}$ &    \textbf{\# iter} &    \textbf{\#visited edges($\times10^3$)}       & $\lVert\q_{\text{offset}}\rVert_{1} $ &   \textbf{\# iter} &    \textbf{\#visited edges($\times10^3$)}
				   & \textbf{L1 norm error}   \\ \hline
		\midrule
		$1$     &    $1$   &    $116$& $3,910,864$   &      $2.60\times10^{-9}$     &     $2$    & $2,145$&     $2.60\times10^{-9}$      &     $1$      &   $25$     &  $2.84\times10^{-8}$\\
		$10$   &    $1$    &    $116$&  $3,910,863$&    $1.51\times10^{-7}$       &      $14$ &  $405,717$&     $1.51\times10^{-7}$      &    $1$       &     $147$    & $3.42\times10^{-7}$\\
		$10^2$&    $1$   &     $116$&   $3,910,858$&    $2.19\times10^{-6}$       &      $26$    & $839,137$&     $2.19\times10^{-6}$      &   $1$        &    $788$ &     $1.77\times10^{-6}$\\
		$10^3$&    $1$    &     $116$&  $3,910,808$&    $2.31\times10^{-5}$       &      $35$  &  $1,169,546$&    $2.31\times10^{-5}$       &   $1$        &      $4,098$ &  $1.64\times10^{-5}$ \\
		$10^4$&    $1$    &    $116$&   $3,910,300$&     $2.30\times10^{-4}$      &      $47$    & $1,604,965$&     $2.30\times10^{-4}$      &  $2$         &    $44,960$  &  $1.11\times10^{-4}$  \\
		$10^5$&    $1$    &     $116$&  $3,905,224$&     $2.05\times10^{-3}$      &      $61$    &  $2,104,446$&      $2.05\times10^{-3}$     &    $4$       &     $130,470$  &  $7.51\times10^{-4}$ \\
		\bottomrule
	\end{tabular}
	\end{threeparttable}
	\vspace{0.5mm}
\end{table*}

\textbf{Fast convergence.}
From Theorem~\ref{theorem:time_dynamic}, \methodD and \methodDA share the same upper bound $O(m)$ for the number of visited edges per iteration with their baseline method \methodC~\cite{YoonJK17}.
In practice, \methodD and \methodDA visit only small portion of edges since the starting vector $\q_{\text{offset}}$ of \methodD and \methodDA has a small $L1$ length:
$\q_{\text{offset}} = (1-c)\DAT\r_{\text{old}}$, and thus $\lVert\q_{\text{offset}}\rVert_{1} \leq (1-c)\lVert(\DA)^\top\rVert_{1}$ with a unit RWR score vector $\r_{\text{old}}$;
then, with small update $\Delta G$, ($\DA)^\top$ is a sparse matrix with small $L1$ length and leads to a small value of $\lVert\q_{\text{offset}}\rVert_{1}$.
When $(\NA + \DA)$ is multiplied with $\q_{\text{offset}}$, only small number of edges in $G + \Delta G$ would be visited.
Table~\ref{visited-edges} shows the $L1$ length of the starting vector ($\lVert\q_{\text{offset}}\rVert_{1}$), the total number of iterations, and the total number of visited edges of \methodC, \methodD, and \methodDA varying the number of deleted edges from the LiveJournal dataset.
The error tolerance $\epsilon$ is set to $10^{-9}$,$10^{-9}$, and $5\times10^{-3}$ for \methodC, \methodD, and \methodDA, respectively.
{\methodD and \methodDA have smaller size of the starting vector $\q_{\text{offset}}$ than \methodC, resulting in fewer numbers of iterations and visited edges.}
Considering the total number of edges ($m$) of the LiveJournal dataset is $34,681,189$, \methodD and \methodDA visit only small portion of edges in the graph thus converge much faster than \methodC does.
\methodDA converges faster than \methodD by trading off accuracy.
{As size of $\Delta G$ increases, numbers of iterations and visited edges, and $L1$ errors all increase.
The reason is analyzed theoretically in Section~\ref{subsec:deltaG}.}

According to Theorem~\ref{theorem:time_dynamic}, error tolerance $\epsilon$ determines the computation cost of \methodDA.
With error tolerance $\epsilon$ and restart probability $c$, we show that error bound of \methodDA is $O(\frac{\epsilon}{c})$ in the following theorem.
\vspace{-2mm}
\begin{theorem}[Error bound of \methodDA]
	\label{theorem:error_dynamic}
	When \methodDA converges under error tolerance $\epsilon$, error bound of RWR score vector $\r_{\text{new}}$ computed by \methodDA is $O(\frac{\epsilon}{c})$.
	\vspace{-2mm}
	\begin{proof}
		When \methodDA iterates until $(k-1)$th iteration, error bound is presented as follows:
		\small
		\begin{align*}
			O(error) &= \lVert\sum_{i=k}^{\infty}x_{\text{offset}}^{(i)}\rVert_{1} = \lVert\sum_{i=k}^{\infty}((1-c)\NBT)^{i}\q_{\text{offset}}\rVert_{1}\\
			&= \lVert\sum_{i=k}^{\infty}((1-c)\NBT)^{i}(1-c)(\NBT-\NAT)\r_{\text{old}}\rVert_{1}\\
			&\leq \sum_{i=k}^{\infty}(1-c)^{i+1}\lVert(\NBT)^{i}\rVert_{1}\lVert\NBT-\NAT\rVert_{1}\lVert\r_{\text{old}}\rVert_{1}\\
			&\leq \sum_{i=k}^{\infty}2(1-c)^{i+1} = \frac{2}{c}(1-c)^{k+1}
		\end{align*}
		\normalsize
		From the proof of Theorem~\ref{theorem:time_dynamic}, $k=\log_{(1-c)}(\frac{\epsilon}{2(1-c)})$. Then error is bounded as:
		\small
		\begin{align*}
			O(error) = \frac{2}{c}\frac{\epsilon}{2(1-c)}(1-c) = \frac{\epsilon}{c}
		\end{align*}
		\normalsize
	\end{proof}
	\vspace{-2.5mm}
\end{theorem}
\vspace{-2mm}
\noindent From Theorem~\ref{theorem:time_dynamic} and Theorem~\ref{theorem:error_dynamic}, \methodDA trades off the running time and accuracy using $\epsilon$.
We show the effects of $\epsilon$ on the experimental performance of \methodDA in Section~\ref{sec:exp-epsilon}.

\subsection{Effects of $\Delta G$}
\label{subsec:deltaG}
Including our model, propagation-based methods~\cite{ohsaka2015efficient,zhang2016approximate} for dynamic RWR computation have sporadically observed long running time which is considerably longer than the average in real-world graphs.
Previous works~\cite{ohsaka2015efficient} detect the fact but do not provide any further investigation on reasons.
Based on \methodDA, we analyze the root cause of the long running time occasionally happened in the propagation models.

From the proof of Theorem~\ref{theorem:time_dynamic}, the running time of \methodDA is determined by the $L1$ length of its seed vector $\q_{\text{offset}}$.
%Note that \methodDA sets $\q_{\text{offset}}$ as a seed vector for \methodDA (Algorithm~\ref{alg:method:dynamic} line 1).
When $\mat{D}$ is a diagonal matrix where $\mat{D}_{ii} = \sum_{j=1}^{n}|\DA_{ij}|$,
$\DNA=\mat{D}^{-1}\DA$ is a row-normalized matrix and $\DNAT$ is a column stochastic matrix.
Then $\q_{\text{offset}}$ and its $L1$ length are presented as follows:
\small
\begin{align*}
	\q_{\text{offset}} &= (1-c)\DAT\r_{\text{old}}\\
	&=(1-c)(\mat{D}\mat{D}^{-1}\DA)^\top\r_{\text{old}}\\
	&=(1-c)(\mat{D}\DNA)^\top\r_{\text{old}}\\
	&=(1-c)\DNAT(\mat{D}\r_{\text{old}}) \\
	\lVert\q_{\text{offset}}\rVert_{1} &= (1-c)\lVert\DNAT(\mat{D}\r_{\text{old}})\rVert_{1}\\
	%&\leq (1-c)\lVert\DNAT\rVert_{1}\lVert\mat{D}\r_{\text{old}}\rVert_{1}\\
	&\leq (1-c)\lVert\mat{D}\r_{\text{old}}\rVert_{1}
\end{align*}
\normalsize
Then $\lVert\mat{D}\r_{\text{old}}\rVert_{1}$ is a decisive factor for running time.
When edges are inserted to node $i$ or deleted from node $i$, $i$th row in $\NB$ is updated from $i$th row in $\NA$;
then $i$th row in $\DA = \NB - \NA$ has nonzero values;
finally, $(i,i)$th element in $\mat{D}$ has a nonzero value.
In summary, $\mat{D}$ is a sparse diagonal matrix which has nonzero values at $(i,i)$th element when node $i$ is modified by a graph modification $\Delta G$.
Then, there are two main components in determining the value of $\lVert\mat{D}\r_{\text{old}}\rVert_{1}$:
1) how many nodes are modified (i.e. the number of nonzeros in $\mat{D}$),
2) which nodes are modified (i.e. the location of nonzeros in $\mat{D}$).

\subsubsection{Size of $\Delta G$}
\label{sec:method:size}
When there are many nodes affected by $\Delta G$, many nonzero values are located in $\mat{D}$'s diagonal and multiplied with $\r_{\text{old}}$.
This leads to a high value of $\lVert\mat{D}\r_{\text{old}}\rVert_{1}$.
In other words, $\lVert\mat{D}\r_{\text{old}}\rVert_{1}$ is determined by the size of $\Delta G$.
Intuitively, when the scope of a graph modification gets larger, the computation time for updating RWR takes longer time.
As shown in Table~\ref{visited-edges}, as $\Delta G$ increases, $\lVert\q_{\text{offset}}\rVert_{1}$ increases in \methodD and \methodDA, then total numbers of iterations and visited edges until convergence also increase.
$L1$ error also increases since the error bound is also determined by $\lVert\q_{\text{offset}}\rVert_{1}$ as shown in the proof of Theorem~\ref{theorem:error_dynamic}.
In Section~\ref{sec:exp-size}, we show the effects of size of $\Delta G$ on performance of \methodDA in real-world graphs.

\subsubsection{Location of $\Delta G$}
\label{sec:method:location}
When the number of nonzeros is fixed,
the location of nonzeros in $\mat{D}$ is a crucial factor for determining the value of $\lVert\mat{D}\r_{\text{old}}\rVert_{1}$.
When nonzeros in $\mat{D}$ are multiplied with high scores in $\r_{\text{old}}$, the product becomes large.
Otherwise, when nonzeros in $\mat{D}$ are multiplied with low scores in $\r_{\text{old}}$, the product becomes small.
In other words, location of $\Delta G$ determines the running time of \methodDA.
When $\Delta G$ appears around high RWR score nodes, running time skyrockets.
On the other hand, when $\Delta G$ appears around low score nodes, \methodDA converges quickly.
Note that most real-world graphs follow power-law degree distribution~\cite{faloutsos1999power} with few nodes having high RWR scores and majority of nodes having low scores.
Thus $\Delta G$ is less likely to happen around high RWR score nodes.
This leads to sporadic occurrence of long running time in propagation models.
In Section~\ref{sec:exp-location}, we show the effects of location of $\Delta G$ on running time of \methodDA in real-world graphs.

\subsection{Discussions}
\label{subsec:discussion}
In addition to edge insertion/deletion, \methodD (and \methodDA) easily handles node insertion/deletion.
We also show how \methodD handles dead-end nodes efficiently. % based on the intuition of \methodC.

\subsubsection{Node insertion/deletion}
\methodD easily handles both node insertion and deletion.
When a node is inserted with its edges, \methodD adds one column and one row, respectively, to the previous $(n\times n)$ matrix $\NA$ with all zero values.
In $(n+1)$th row of the updated matrix $\NB$, edge distribution of the new node is stored.
Likewise, \methodD adds one row to the $(n\times1)$ vector $\r_{\text{old}}$ with a zero value and stores an RWR score of the new node in $(n+1)$th row of $\r_{\text{new}}$.
On the other hand, when a node is deleted from a given graph, the corresponding row in $\NB$ is simply set to all zero values to express the deletion.
The remaining process is the same as that of edge insertion/deletion as described in Algorithm~\ref{alg:method:dynamic}.

\subsubsection{Dead-end}
Dead-ends which do not have any out-edges cause scores to leak out.
Without handling dead-ends, total sum of RWR scores across a given graph would be less than $1$.
One common way~\cite{zhang2016approximate} to tackle the leakage problem is inserting edges from dead-end nodes to a seed node.
However, inserting edges for every dead-end node leads to explosive computation time
as the given graph gets larger proportional to the number of dead-ends.
Thus RWR experiments have been frequently conducted without handling dead-ends~\cite{ohsaka2015efficient,zhang2016approximate}.
Rather than inserting new edges, we handle the dead-end problem in an efficient way.
In \methodC, an initial score $c$ is generated at the seed node, and propagated across the graph.
The previous dead-end handling hands over scores inserted into dead-ends to the seed node.
In this sense, we collect {whole scores handed over to the seed node from dead-ends} and call it $d_{\text{total}}$.
We do not need to propagate $d_{\text{total}}$ from the seed node since we already have the result of propagating score $c$ from the seed node.
{
$\r_{\text{temp}} = c\sum_{i=0}^{\infty}((1-c)\NBT)^{i}\q$ denotes a result score vector with the initial score $c$ without any handling for dead-ends.
Then we scale $\r_{\text{temp}}$ with $\frac{d_{\text{total}}}{c}$ and get $d_{\text{total}}\sum_{i=0}^{\infty}((1-c)\NBT)^{i}\q$ which is the result of propagation with initial score $d_{\text{total}}$.
}
Furthermore, we do not need to collect $d_{\text{total}}$ since we know that $L1$ length of an RWR score vector is $1$.
Finally, we handle dead-ends by scaling $\r_{\text{temp}}$ to $\frac{\r_{\text{temp}}}{\lVert\r_{\text{temp}}\rVert_{1}}$ to get a final RWR score vector without requiring extra computation time.
We prove the exactness of the dead-end handling described above in Theorem~\ref{theorem:dead-end}.

\begin{theorem}[Exactness of Dead-end handling]
	\label{theorem:dead-end}
	When $\r_{\text{temp}}$ denotes a result score vector in \methodD without any handling for dead-ends,
	scaling $\r_{\text{temp}}$ to $\frac{\r_{\text{temp}}}{\lVert\r_{\text{temp}}\rVert_{1}}$ results in an RWR score vector $\r_{\text{final}}$ with dead-end handling.
	\vspace{-2mm}
	\begin{proof}
		Let $d_{\text{total}_1}$ be the whole score handed over to the seed node from dead-ends after propagating the initial score $c$ from the seed.
		Then $d_{\text{total}_1}$ is propagated across the graph to fill the leaked scores.
		We do not need to calculate the propagation of $d_{\text{total}_1}$ from the seed node:
		$\r_{\text{temp}}=c\sum_{i=0}^{\infty}((1-c)\NBT)^{i}\q$, then scaling $\r_{\text{temp}}$ with $(d_{\text{total}_1}/c)$ results in $d_{\text{total}_1}\sum_{i=0}^{\infty}((1-c)\NBT)^{i}\q$
		{which is the result of propagating score $d_{\text{total}_1}$ from the seed}.
		At this time, the propagation of $d_{\text{total}_1}$ is leaked out again from the dead-ends.
		Repeatedly, we collect the leaked scores $(d_{\text{total}_2}, d_{\text{total}_3}, \cdots)$ which are inserted into the dead-ends, and propagate them from the seed.
		However, we do not need to collect the leaked scores $d_{\text{total}_1}, d_{\text{total}_2}, \cdots$ and propagate them from the seed again and again.
		We know that
		\vspace{-1.5mm}
		\small
		\begin{align*}
		\lVert\r_{\text{temp}}\rVert_{1} = c\lVert\sum_{i=0}^{\infty}((1-c)\NBT)^{i}\q\rVert_{1}\\
		\lVert\sum_{i=0}^{\infty}((1-c)\NBT)^{i}\q\rVert_{1} =\frac{\lVert\r_{\text{temp}}\rVert_{1}}{c}
		\end{align*}
		\normalsize
		Then the final RWR vector $\r_{\text{final}}$ with dead-end handling is presented as follows:
		\small
		\begin{align*}
		\r_{\text{final}} &= c\sum_{i=0}^{\infty}((1-c)\NBT)^{i}\q + d_{\text{total}_1}\sum_{i=0}^{\infty}((1-c)\NBT)^{i}\q \\
		&+ d_{\text{total}_2}\sum_{i=0}^{\infty}((1-c)\NBT)^{i}\q + \cdots\\
		 &=(c+d_{\text{total}_1}+d_{\text{total}_2}+d_{\text{total}_3}+\cdots)\sum_{i=0}^{\infty}((1-c)\NBT)^{i}\q\\
		\lVert\r_{\text{final}}\rVert_{1} &=(c+d_{\text{total}_1}+d_{\text{total}_2}+d_{\text{total}_3}+\cdots)\lVert\sum_{i=0}^{\infty}((1-c)\NBT)^{i}\q\rVert_{1} = 1
		\end{align*}
		\normalsize
		Then $(c+d_{\text{total}_1}+d_{\text{total}_2}+d_{\text{total}_3}+\cdots)$ is expressed as follows:
		\small
		\begin{align*}
		c+d_{\text{total}_1}+d_{\text{total}_2}+d_{\text{total}_3}+\cdots &= 1/\lVert\sum_{i=0}^{\infty}((1-c)\NBT)^{i}\q\rVert_{1} =\frac{c}{\lVert\r_{\text{temp}}\rVert_{1}}
		\end{align*}
		\normalsize
		Thus we get $\r_{\text{final}}$ by scaling $\r_{\text{temp}}$ with $1/\lVert\r_{\text{temp}}\rVert_{1}$ as follows:
		\small
		\begin{align*}
		\vspace{-1.5mm}
		\r_{\text{final}} &=(c+d_{\text{total}_1}+d_{\text{total}_2}+d_{\text{total}_3}+\cdots)\sum_{i=0}^{\infty}((1-c)\NBT)^{i}\q\\
		\vspace{-5mm}
		&= \frac{c}{\lVert\r_{\text{temp}}\rVert_{1}}\sum_{i=0}^{\infty}((1-c)\NBT)^{i}\q\\
		&=\frac{1}{\lVert\r_{\text{temp}}\rVert_{1}}\r_{\text{temp}}
		\end{align*}
		\normalsize
		\vspace{-10mm}
	\end{proof}
\end{theorem}
%With help of this efficient dead-end handling, \methodD and \methodDA earn high accuracy avoiding explosive extra computation time of the conventional dead-end handling.	

\vspace{5pt}
\section{Experiments}
\label{sec:experiments}
\begin{table}[!t]
	\begin{threeparttable}[t]
		\centering
		\small
		\caption{Dataset statistics}
		\begin{tabular}{C{15mm} | R{12mm} R{15mm} R{13mm} R{13mm}}\hline
			\toprule
			\textbf{Dataset} & \textbf{Nodes} & \textbf{Edges} & \textbf{Direction} &\textbf{Error tolerance (\methodDA)} \\
			\midrule
			WikiLink\tnote{1}&    12,150,976&    378,142,420&	Directed&$	10^{-2}$ \\
			Orkut\tnote{1}&    3,072,441&    117,185,083&	Undirected &$	 5\times10^{-3}$\\
			LiveJournal\tnote{1}&    3,997,962&    34,681,189&	Undirected &$	 5\times10^{-3}$ \\
			Berkstan\tnote{1}&    685,230&    7,600,595&	Directed &	$10^{-4}$ \\
			DBLP\tnote{1}&    317,080&    1,049,866&	Undirected  &	$10^{-4}$\\
			Slashdot\tnote{1}&    82,144&    549,202&	Directed &	$10^{-4}$\\
			\bottomrule
		\end{tabular}
		\label{tab:dataset}
		\begin{tablenotes}
			%\item[1] {{http://konect.uni-koblenz.de/networks/}}
			\item[1] {{http://snap.stanford.edu/data/}}
			\vspace{1mm}
		\end{tablenotes}
	\end{threeparttable}
\end{table}

\begin{figure*}[!t]
	\centering
	\vspace{-2mm}
	\includegraphics[width=.28\linewidth]{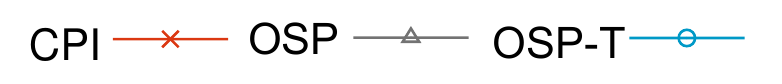}\\
	\vspace{-1mm}
	\hspace{-2mm}
	\subfigure[DBLP]
	{
		\includegraphics[width=.24\linewidth]{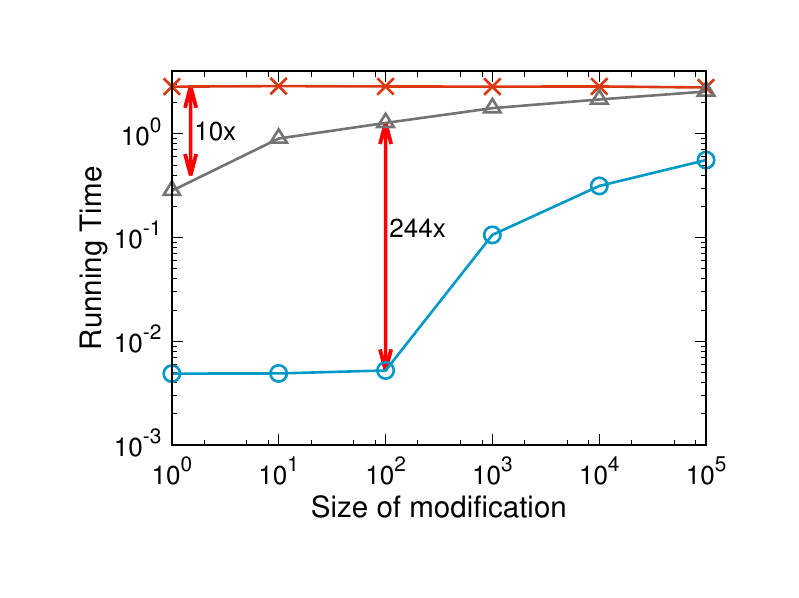}
	}
	\subfigure[Berkstan]
	{
		\includegraphics[width=.24\linewidth]{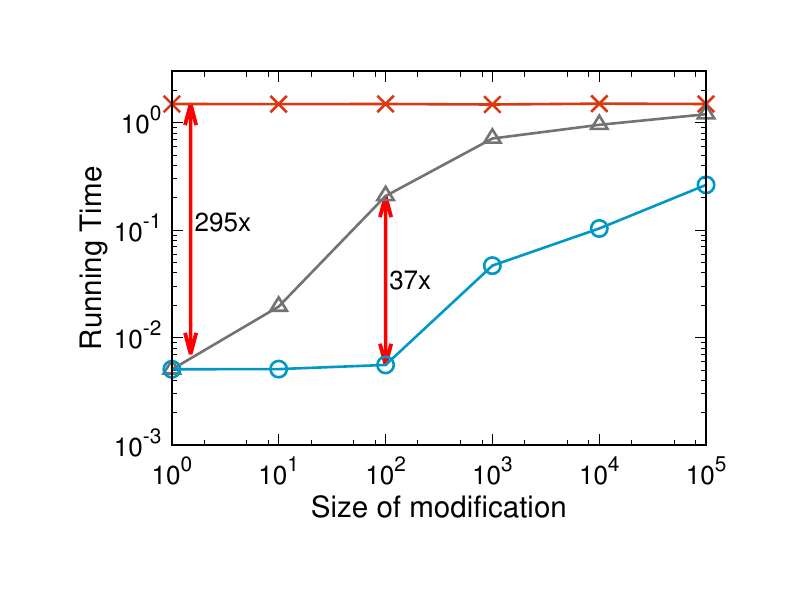}
	}
	\subfigure[LiveJournal]
	{
	
		\includegraphics[width=.24\linewidth]{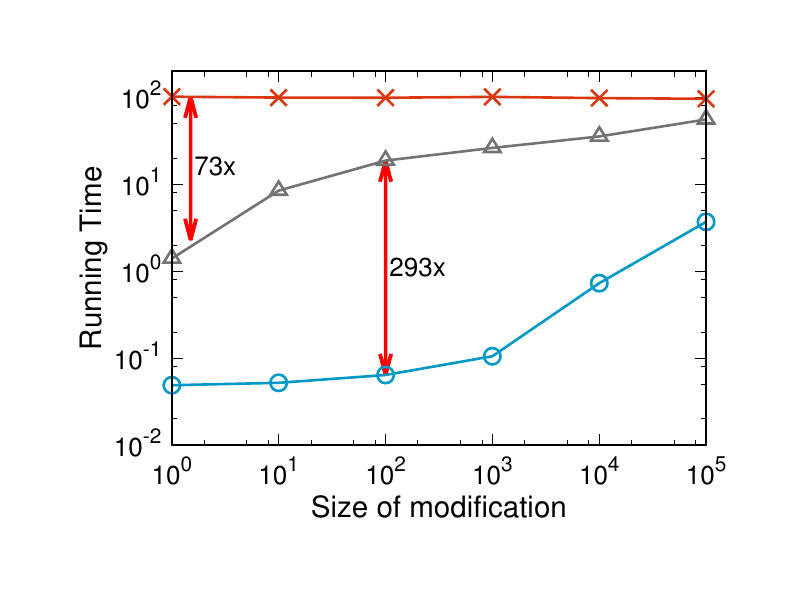}
	}
	\subfigure[Orkut]
	{
		\includegraphics[width=.24\linewidth]{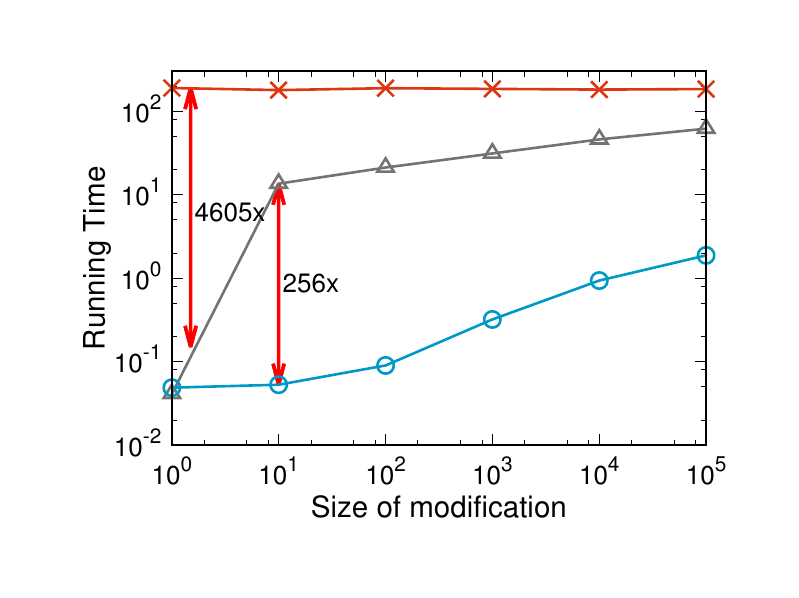}
	}
	\caption
	{
		Performance of \methodD and \methodDA:
		\methodD computes the exact RWR faster than \methodC on dynamic graphs.
		%As size of modification gets larger, computation time of \methodD is converged to \methodC.
		\methodDA achieves faster speed than \methodD while sacrificing accuracy using higher error tolerance $\epsilon$.
		$\epsilon$ is set to $10^{-9}$ for \methodC and \methodD, and set to higher values for \methodDA as described in Table~\ref{tab:dataset}.
		Experiments for accuracy of \methodDA are presented in Section~\ref{sec:exp-size}.
	}
	\label{fig:exact}
\end{figure*}

In this section, we experimentally evaluate the performance of \methodD and \methodDA compared to other dynamic RWR methods.
We aim to answer the following questions:
\begin{itemize}
	{\bfseries {\item Q1 Performance of \methodD.}}
	How much does \methodD improve performance for dynamic RWR computation from the baseline static method \methodC? (Section~\ref{sec:exp-exact})
	{\bfseries {\item Q2 Performance of \methodDA.}}
	How much does \methodDA enhance computation efficiency, accuracy and scalability compared with its competitors? (Section~\ref{sec:exp-dynamic})
	{\bfseries {\item Q3 Effects of \textnormal{$\epsilon$}, error tolerance.}}
	How does the error tolerance $\epsilon$ affect the accuracy and the speed of \methodDA? (Section~\ref{sec:exp-epsilon})
	{\bfseries {\item Q4 Effects of \textnormal{$\Delta G$}, a graph modification.}}
	How does the size of $\Delta G$ affect the performance of \methodDA? (Section~\ref{sec:exp-size})
	How does the location of $\Delta G$ in the given graph $G$ affect the performance of \methodDA? (Section~\ref{sec:exp-location})
\end{itemize}

\subsection{Setup}
\label{sec:exp-setup}
\subsubsection{Datasets}
\label{sec:exp-dataset}
We use 6 real-world graphs to evaluate the effectiveness and efficiency of our methods.
The datasets and their statistics are summarized in Table~\ref{tab:dataset}.
Among them, Orkut, LiveJournal, and Slashdot are social networks, whereas DBLP is a collaboration network, and WikiLink and Berkstan are hyperlink networks.
%We divide an undirected edge into two directed edges in undirected graphs for implementation.

\subsubsection{Environment}
\label{sec:exp-environment}
All experiments are conducted on a workstation with a single core Intel(R) Xeon(R) CPU E5-2630 @ 2.2GHz and 512GB memory.
We compare \methodD with its baseline static method \methodC, and
compare \methodDA with two state-of-the-art approximate methods for dynamic RWR, TrackingPPR~\cite{ohsaka2015efficient} and LazyForward~\cite{zhang2016approximate}, all of which are described in Section~\ref{sec:related_works}.
All these methods including \methodD and \methodDA are implemented in C++.
We set the restart probability $c$ to $0.15$ as in the previous works~\cite{ohsaka2015efficient,YoonJK17}.
For each dataset, we measure the average value for $30$ random seed nodes.
\methodC~\cite{YoonJK17} is used to provide the exact RWR values in all experiments.

\subsection{Performance of \methodD}
\label{sec:exp-exact}

We evaluate performance of \methodD by measuring computation time for tracking RWR exactly on a dynamic graph $G$ varying the size of $\Delta G$.
We set the initial graph G as a graph with all its edges and modify $G$ by deleting edges.
The size of $\Delta G$ varies from one edge to $10^5$ edges.
For space efficiency, we show the results on Orkut, LiveJournal, Berkstan, and DBLP; results on other graphs are similar.
Error tolerance for \methodC and \methodD is set to $10^{-9}$ for all datasets, and error tolerance for \methodDA on each dataset is described in Table 4.
From Figure~\ref{fig:exact}, \methodD tracks the exact RWR on dynamic graphs up to $4605\times$ faster than \methodC with the help of small size of the starting vector $\q_{\text{offset}}$.
\methodDA trades off accuracy against speed using higher error tolerance $\epsilon$, thus results in superior speed than \methodC and \methodD.
Note that \methodC and \methodD compute the exact RWR scores while \methodDA results in the approximate RWR scores on dynamic graphs.
As size of $\Delta G$ becomes larger, computation time of \methodD and \methodDA increases while \methodC maintains similar computation time.
The effects of size of $\Delta G$ on computation time of \methodD and \methodDA are discussed concretely in Section~\ref{sec:exp-size}.
%As expected, when whole graph is modified, computation time for \methodD would be converged to computation time of \methodC.
%As size of $\Delta G$ becomes larger, computation time of \methodC decreases slightly due to the decreased size of the graph with the smaller adjacency matrix.

\subsection{Performance of \methodDA}
\label{sec:exp-dynamic}
From each dataset, we generate a uniformly random edge stream and divide the stream into two parts.
We extract $10$ snapshots from the second part of the edge stream.
At first, we initialize each graph with the first part of the stream, and then update the graph for each new snapshot arrival.
%depending on the algorithm we are using for each new snapshot arrival.
At the end of the updates, we compare the performance of each algorithm.
\subsubsection{Computational Efficiency}
\label{sec:exp-efficiency}
\begin{figure*}[!t]
	\centering
	\vspace{-2mm}
	\includegraphics[width=.85\linewidth]{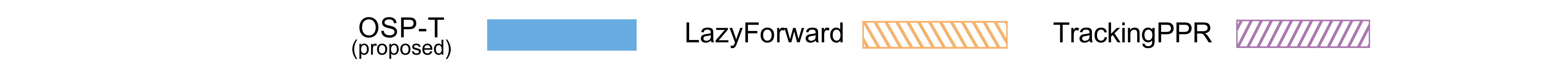}\\
	\vspace{-2.5mm}
	\hspace{-5mm}
	\subfigure[Running time]
	{
		\label{fig:perf:time}
		\includegraphics[width=.32\linewidth]{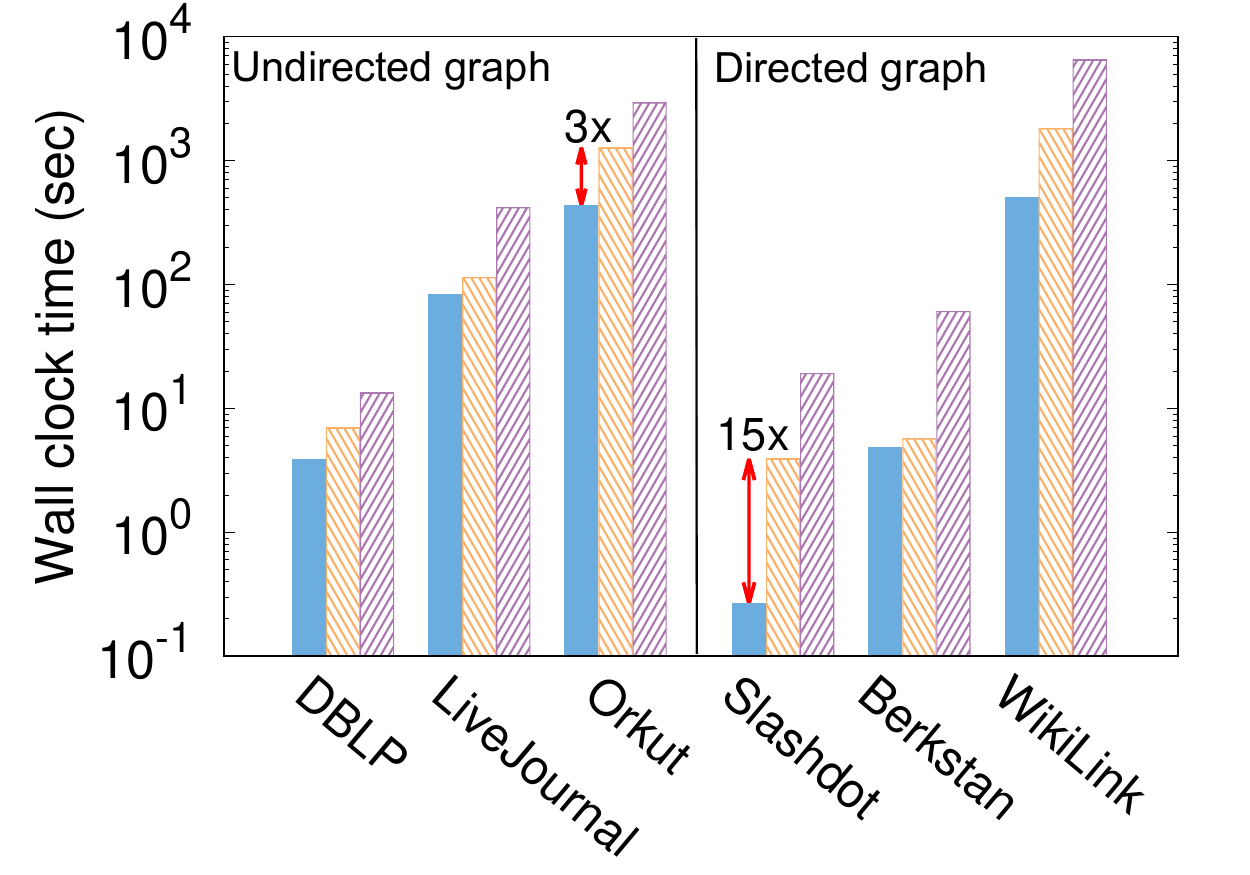}
	}
	\subfigure[Accuracy on L1 norm of error]
	{
		\label{fig:perf:l1_error}
		\includegraphics[width=.32\linewidth]{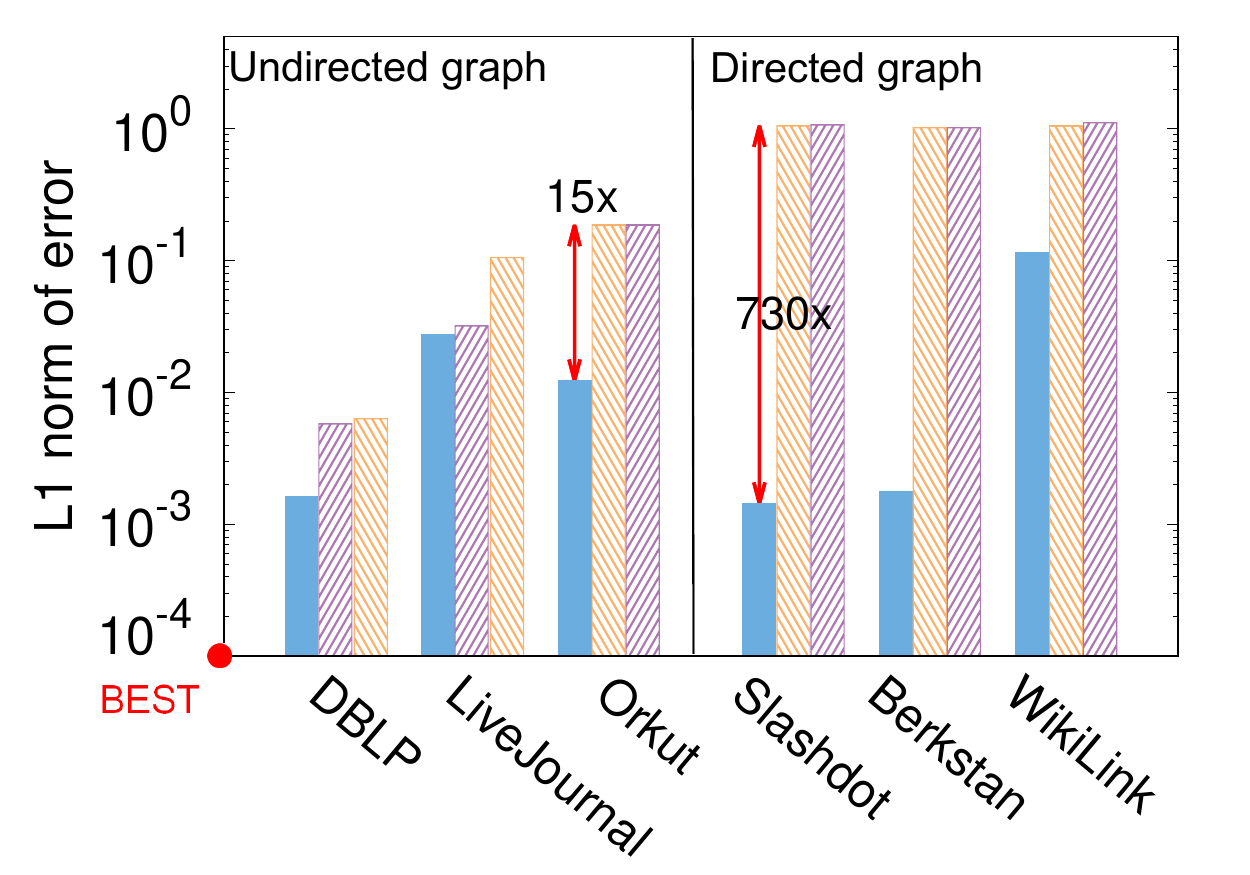}
	}
	\subfigure[Accuracy on Rank]
	{
		\label{fig:perf:rank_error}
		\includegraphics[width=.32\linewidth]{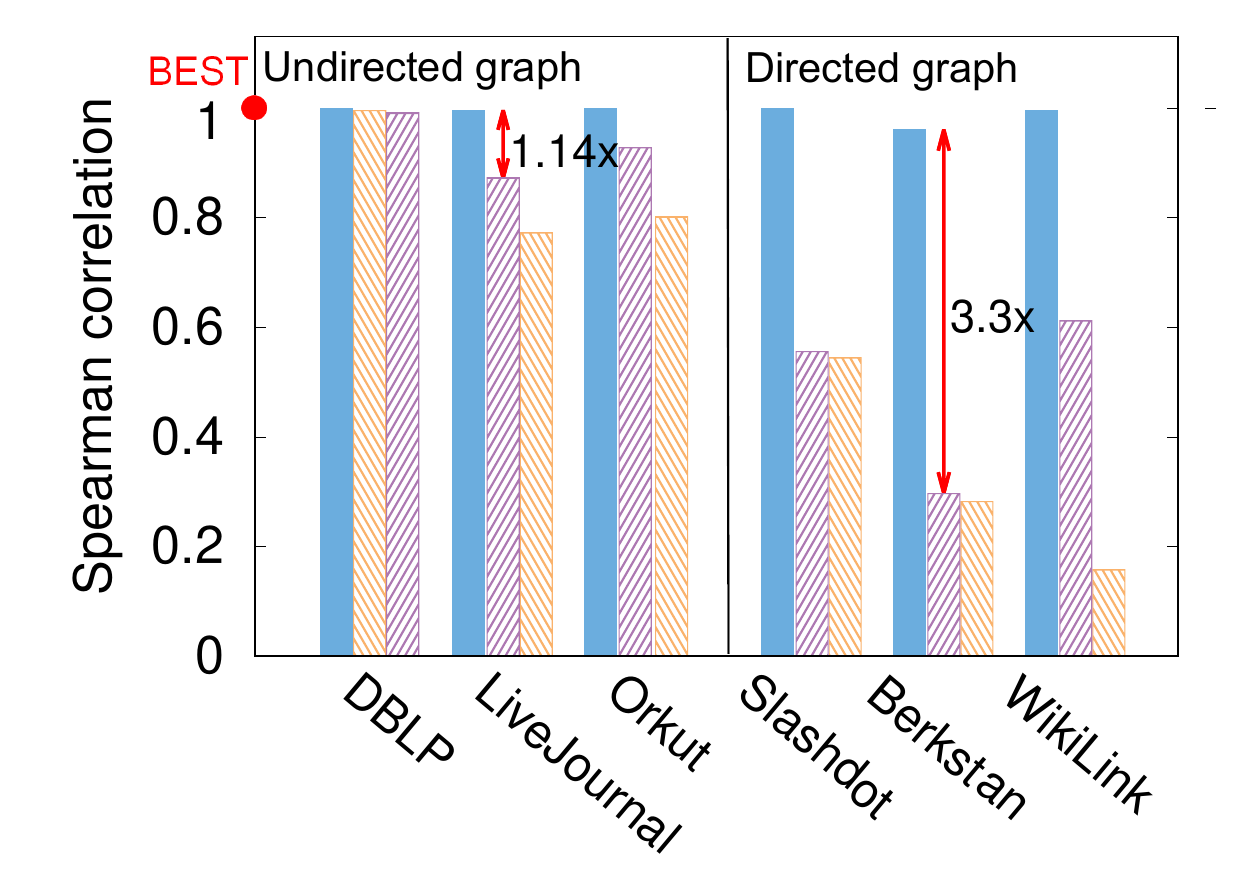}
	}
	\caption
	{ Performance of \methodDA: (a) compares the running time among dynamic RWR methods;
		(b) and (c) compare the L1 norm of error and the rank accuracy of RWR scores of \methodDA and other methods with those of the exact RWR scores, respectively.
		While other methods show different performance on directed/undirected graphs, \methodDA maintains superior performance on overall graphs.
		%(a) \methodDA tracks RWR scores faster than other competitors over all dynamic datasets.
		%Only \methodD successfully preprocesses billion-scale graphs.
		%(b) \methodDA provides the highest accuracy for RWR among all tested methods in terms of L1 norm error.
		%(c) \methodDA shows the highest rank accuracy among competitors.
		%The higher the Spearman correlation, the higher is the rank accuracy.
		%Details on these experiments are presented in Section~\ref{sec:experiments}.
	}
	\label{fig:perf}
\end{figure*}

We evaluate the computational efficiency of \methodDA in terms of running time when error tolerance is given.
Error tolerance used in TrackingPPR and LazyForward means maximum permissible $L1$ error per node, while error tolerance in \methodDA indicates maximum permissible $L1$ error per RWR score vector.
For TrackingPPR and LazyForward, we set error tolerance to $10^{-8}$ across all datasets.
For \methodDA, we set error tolerance close to $10^{-8}\times(\# nodes)$ for each graph, respectively, to give the same error tolerance effect with the competitors.
Error tolerance $\epsilon$ for each dataset is described in Table~\ref{tab:dataset}.
The wall-clock running time is shown in Figure~\ref{fig:perf:time}.
\methodDA runs faster than other methods by up to $15\times$ while maintaining higher accuracy as shown in Figures~\ref{fig:perf:l1_error} and~\ref{fig:perf:rank_error}.
This performance difference comes from the different definitions of error tolerance which are described earlier.
In \methodDA, error tolerance works per RWR score vector, and thus convergence condition is checked after all nodes that could be reached in one hop are updated.
On the other hand, error tolerance works per node in TrackingPPR and LazyForward, and thus convergence condition is checked every time a node is updated.

\subsubsection{Accuracy}
\label{sec:exp-accuracy}
After all updates with snapshots, we get an approximate RWR vector for a given graph from each method.
%In many applications of RWR, one common approach is to return the ranking in nodes of RWR vector.
%For example, in Twitter's "Who to Follow" recommendation service, the top-ranked users in RWR will be recommended.
%On the other hand, for the trolling detection, the low-ranked users will be returned.
%Therefore, it is important to measure the ranking error to examine the accuracy of an approximate RWR vector.
We compare $L1$ norm error between an approximate RWR vector and its exact RWR vector.
To measures the rank accuracy, we use Spearman correlation~\cite{artusi2002bravais}. % as in the previous works~\cite{YoonJK17}.
Note that the higher the Spearman correlation, the higher is the rank accuracy.
As shown in Figures~\ref{fig:perf:l1_error} and \ref{fig:perf:rank_error}, \methodDA outperforms other state-of-the-art methods with higher accuracy by up to $730\times$ in $L1$ norm and $3.3\times$ in ranking.
This difference in accuracy comes from the different propagation models used in \methodDA and its competitors.
Gauss-Southwell algorithm and Forward Local Push algorithm used in TrackingPPR and LazyForward, respectively, propagate scores toward the top-ranked node in terms of residual scores,
while \methodC used in \methodDA propagates scores toward the whole nodes which could be reached in one hop.
Note that \methodDA maintains high accuracy on overall graphs, whereas LazyForward and TrackingPPR show different accuracy on directed and undirected graphs;
since Gauss-Southwell algorithm and Forward Local Push algorithm are based on Local Push algorithm~\cite{andersen2006local} which is designed for undirected graphs, they both show considerably lower accuracy on directed graphs than on undirected graphs.

\begin{figure}[!t]
	\centering
	%\vspace{1mm}
	\includegraphics[width=.72\linewidth]{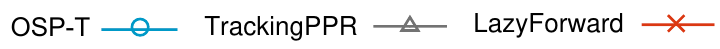}\\
	\hspace{-7mm}
	\vspace{-3mm}
	\subfigure[Insertion on Orkut]
	{
		\includegraphics[width=.5\linewidth]{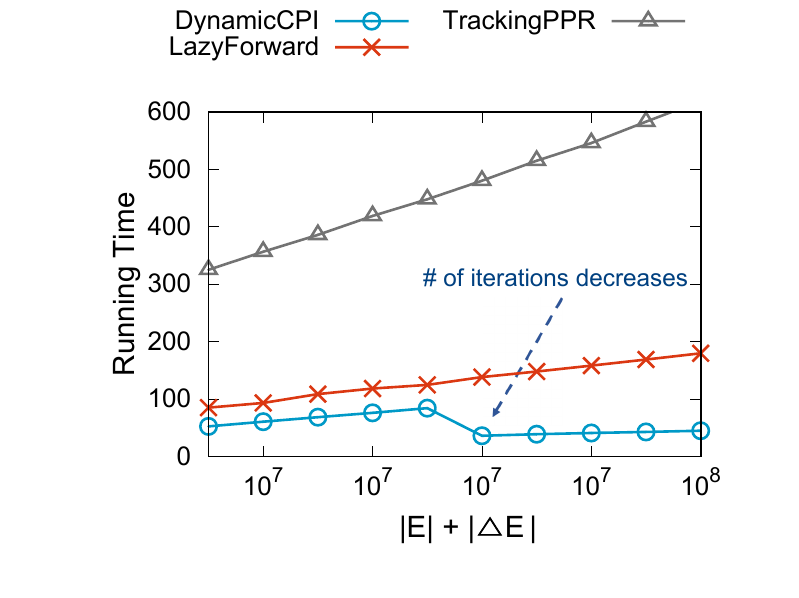} %0.77
		\label{fig:scalability:insert}
	}
	\subfigure[Deletion on Orkut]
	{
		\hspace{-3mm}
		\includegraphics[width=.5\linewidth]{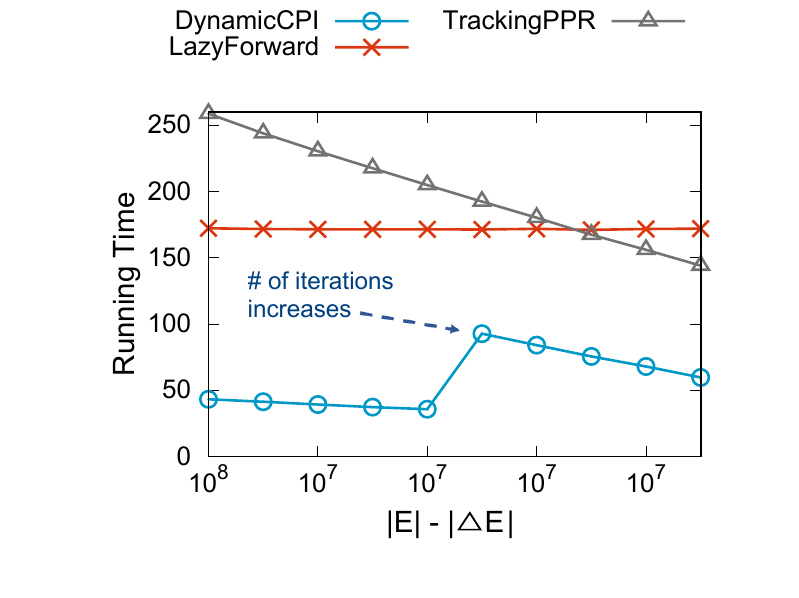} %0.7
		\label{fig:scalability:delete}
	}
	\vspace{3mm}
	\caption
	{
		Scalability:
		\methodDA shows the best scalability among the competitors.
	}
	\label{fig:scalability}
\end{figure}

\subsubsection{Scalability}
\label{sec:exp-scalability}
\begin{figure}[!t]
	\centering
	\vspace{-1mm}
	\includegraphics[width=.72\linewidth]{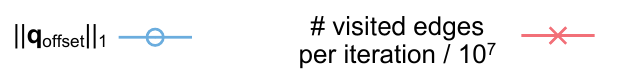}\\
	\vspace{-3mm}
	\hspace{-5mm}
	\subfigure[Insertion on Orkut]
	{
		\includegraphics[width=.5\linewidth]{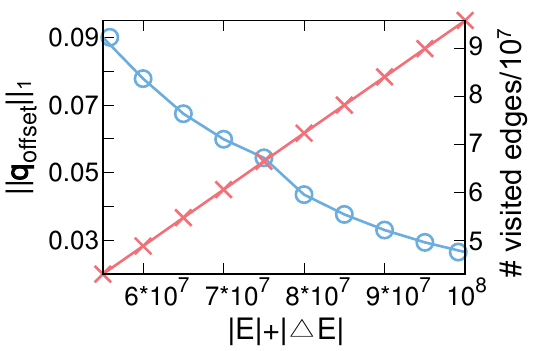}
		\label{fig:scalability2:insert}
	}
	\subfigure[Deletion on Orkut]
	{
		\hspace{-2mm}
		\includegraphics[width=.5\linewidth]{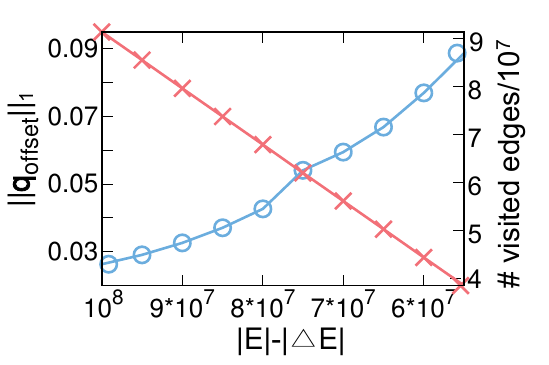}
		\label{fig:scalability2:delete}
	}
	\hspace{-5mm}
	\caption
	{
		Two deciding factors for running time:
		$\lVert\q_{\text{offset}}\rVert_{1}$ and the number of visited edges per iteration show different tendencies when a graph grows/shrinks.
		Note that $|\Delta E| $ is fixed.
	}
	\label{fig:scalability2}
\end{figure}

In this experiment, we estimate scalability of each method by comparing running time when a given graph incrementally grows/shrinks by inserting/deleting $\Delta G$ of a fixed size.
For brevity, we show the result on Orkut; results on other graphs are similar.
$|\Delta G|$ is fixed to $5\times10^6$ edges and the error tolerance for each method is the same as that in Section~\ref{sec:exp-efficiency}.
When the graph incrementally grows, \methodDA shows different tendency compared to other methods.
While all methods take a longer time as the graph grows, \methodDA occasionally shows a sudden drop in time as shown in Figure~\ref{fig:scalability:insert}.
Similarly, \methodDA shows a sudden jump when the given graph incrementally shrinks as shown in Figure~\ref{fig:scalability:delete}.
From the proof of Theorem~\ref{theorem:time_dynamic}, time complexity of \methodDA is $O(m\log_{(1-c)}(\frac{\epsilon}{\lVert\q_{\text{offset}}\rVert_{1}}))$.
The first term $m$ indicates the upper bound of number of edges which could be visited in each iteration.
When a graph grows, the number of visited edges in each iteration would increase as shown in Figure~\ref{fig:scalability2:insert}.
The second term $\log_{(1-c)}(\frac{\epsilon}{\lVert\q_{\text{offset}}\rVert_{1}})$ indicates the number of iterations needed to converge, and is positively correlated with $\lVert\q_{\text{offset}}\rVert_{1}$.
From Section~\ref{subsec:deltaG}, $\lVert\q_{\text{offset}}\rVert_{1}$ is decided by RWR scores of updated nodes:
as RWR scores of updated nodes increase, $\lVert\q_{\text{offset}}\rVert_{1}$ increases.
Assume that $\Delta G$ of a fixed size updates $k$ nodes in the graph.
As the graph grows, the average RWR score of the $k$ nodes decreases since the total number of nodes in the graph increases while the total sum of RWR scores among nodes is always 1.
As a result, as the graph grows, $\lVert\q_{\text{offset}}\rVert_{1}$ becomes smaller as shown in Figure~\ref{fig:scalability2:insert}.
This leads to fewer iterations for convergence.
In Figure~\ref{fig:scalability:insert}, the number of iterations changes from $3$ to $2$ when $|E|+|\Delta E|$ is $8\times10^7$, thus the running time suddenly drops.
When the number of iterations is consistent $(5.5\times10^{7} < |E|+|\Delta E| < 7.5\times10^{7} $ and $8\times10^{7} < |E|+|\Delta E|  < 10^{8})$,
the running time is decided by the first term $m$ thus increasing constantly as the graph grows.
In case of deletion, the opposite process is applied.
As the graph shrinks, the average RWR score of updated nodes increases, thus $\lVert\q_{\text{offset}}\rVert_{1}$ becomes larger as shown in Figure~\ref{fig:scalability2:delete}.
This leads to increased number of iterations for convergence.
The number of iterations changes from $2$ to $3$ when $|E|-|\Delta E|$ is $7.5\times10^7$ with a sudden jump in running time.
When the number of iterations is consistent, the running time decreases constantly as the graph shrinks.
In LazyForward, time complexity of an undirected graph is $O(|\Delta E| + 1/(\bar{d}\delta))$~\cite{zhang2016approximate} where
$\delta$ is the error tolerance per node and $\bar{d}$ is the average degree
~\footnote{In \cite{zhang2016approximate}, the original time complexity is $O(|\Delta E| + 1/\bar{\delta})$ where
$\bar{\delta}$ is a degree-normalized error tolerance per node such that $|\r^{(i)}(t)|/d(t)<\bar{\delta}$ for residuals $r$ with any node $t$ at $i$th iteration.
To consider the effect of changes in degree, we present $\delta$ as $\bar{d}\bar{\delta}$ with the average degree $\bar{d}$.
}
of the graph which increases as the graph grows.
In the time complexity model, running time would decrease as the graph grows, and increase as the graph shrinks.
However, running time constantly increases when the graph grows in Figure~\ref{fig:scalability:insert}
and maintains a constant value when the graph shrinks in Figure~\ref{fig:scalability:delete}.
Thus, the time complexity model fails to explain scalability of LazyForward.
%However, deletion does not follow the time complexity model.
Since TrackingPPR~\cite{ohsaka2015efficient} does not provide a time complexity model for general insertion/deletion, we could not investigate further.
Only \methodDA succeeds in analyzing its scalability based on its time complexity model.

\subsection{Effects of \textnormal{$\epsilon$}, error tolerance}
\label{sec:exp-epsilon}
\begin{figure}[!t]
	\centering
	\vspace{-2mm}
	\includegraphics[width=.9\linewidth]{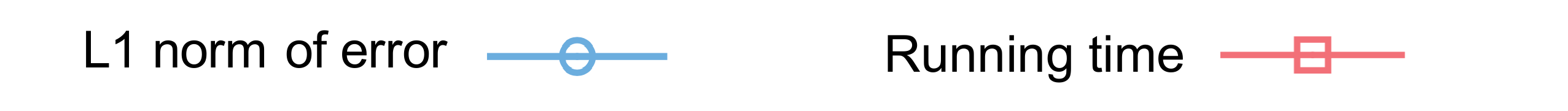}\\
	\vspace{-4mm}
	\hspace{-5mm}
	\subfigure[DBLP]
	{
		\includegraphics[width=.5\linewidth]{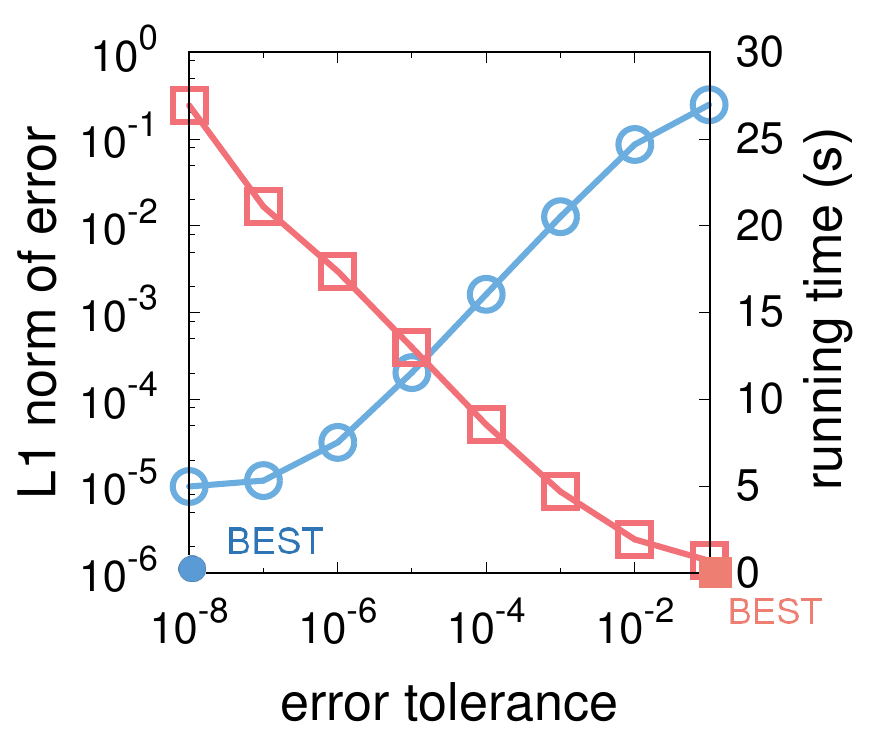}
	}
	\subfigure[Berkstan]
	{
		\includegraphics[width=.5\linewidth]{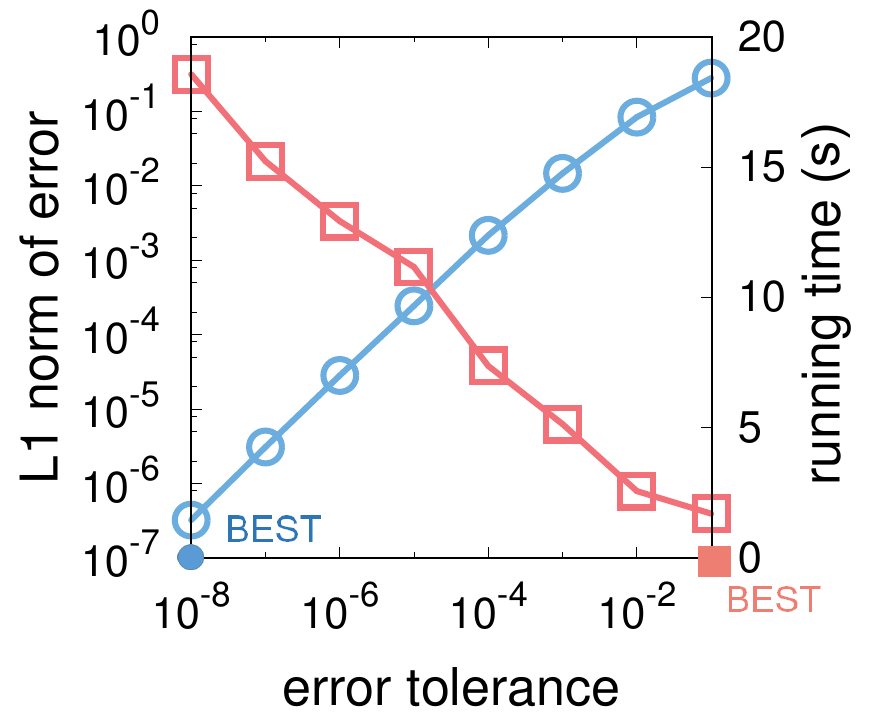}
	}
	\hspace{-5mm}
	\caption
	{
		Effects of error tolerance $\epsilon$ on \methodDA:
		as $\epsilon$ increases, the $L1$ error increases while the running time decreases.
	}
	\label{fig:epsilon}
\end{figure}

To examine the effects of error tolerance $\epsilon$ on the performance of \methodDA, we check the $L1$ error and running time varying $\epsilon$.
We report results on DBLP and Berkstan for brevity; results on other graphs are similar.
As shown in Figure~\ref{fig:epsilon}, as $\epsilon$ increases, the running time of \methodDA decreases and $L1$ error increases across all datasets.
%directly proportional
Note that we theoretically proved O(running time) is proportional to $log(\epsilon)$ in Theorem~\ref{theorem:time_dynamic}, and O($L1$ error) is proportional to $\epsilon$ in Theorem~\ref{theorem:error_dynamic}.
We verify those theorems experimentally in this experiment.
In Figure~\ref{fig:epsilon}, error tolerance and $L1$ error are plotted in log scale while running time is plotted in linear scale.
Relations among them are near linear.
This shows that the theorems describing the relations among $L1$ error, running time and $\epsilon$ work in real-world datasets.

\begin{figure}[!t]
	\centering
	%\vspace{1mm}
	\includegraphics[width=.8\linewidth]{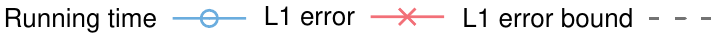}\\
	\vspace{-2mm}
	\hspace{-7mm}
	\subfigure[DBLP]
	{
		\includegraphics[width=.5\linewidth]{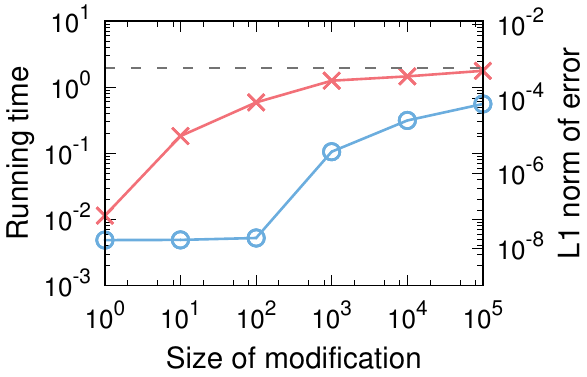}
	}
	\hspace{-2mm}
	\subfigure[LiveJournal]
	{
		\includegraphics[width=.5\linewidth]{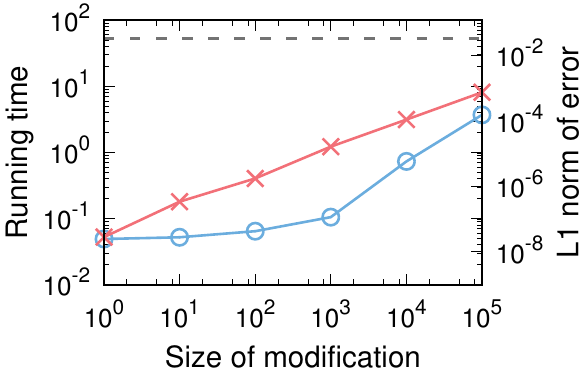}
	}
	\caption
	{
		Effects of size of $\Delta G$ on \methodDA:
		as the size of $\Delta G$ becomes bigger, both computation time and $L1$ error increase.
	}
	\label{fig:delta:size}
\end{figure}

\begin{figure}[!t]
	\centering
	\vspace{-3mm}
	\hspace{-5mm}
	\subfigure[DBLP]
	{
		\includegraphics[width=.5\linewidth]{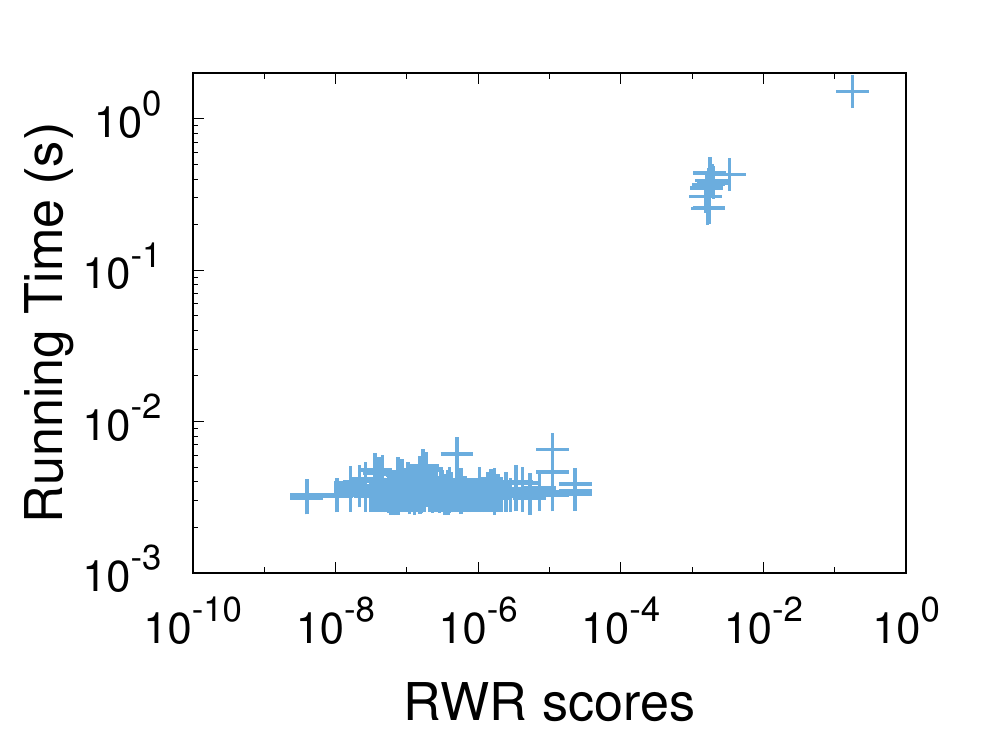}
	}
	\hspace{-2mm}
	\subfigure[Berkstan]
	{
		\includegraphics[width=.5\linewidth]{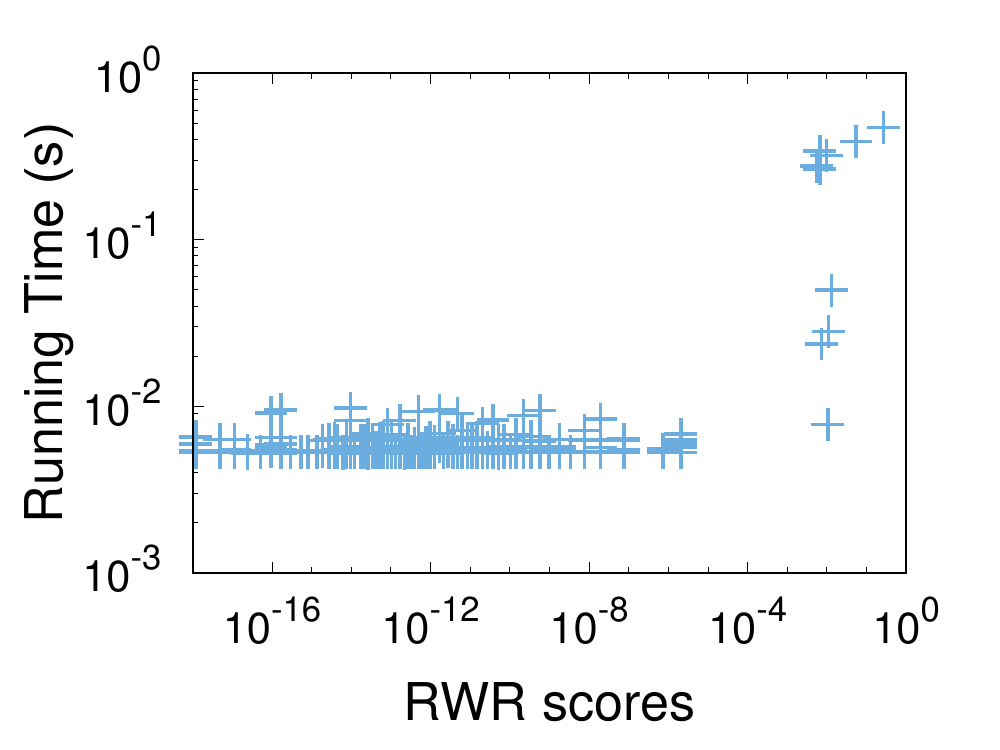}
	}
	\caption
	{
		Effects of location of $\Delta G$:
		when $\Delta G$ happens around high RWR score nodes, \methodDA takes long running time.
		However, this rarely happens since there are only few nodes with high RWR score in a graph.
	}
	\vspace{1mm}
	\label{fig:delta:rwr}
\end{figure}

\subsection{Effects of $\Delta G$, a graph modification}
\label{sec:exp-delta}
%In the following experiments, we show the effectiveness of \textsc{Dynamic}\textsc{CPI-T} for edge deletion, by setting an initial graph $G$ as a graph with all its edges and modifying $G$ by deleting edges.
%Note that edge insertion shows similar results.

\subsubsection{Size of $\Delta G$}
\label{sec:exp-size}
%To show the effects of size of a graph modification $\Delta G$ on the performance of \methodDA, we estimate the running time of \methodDA varying the size of $\Delta G$.
%The size of $\Delta G$ is varying from one edge to $10^5$ edges.
In Figure~\ref{fig:delta:size}, as the size of $\Delta G$ increases, the running time and $L1$ error of \methodDA also increase.
%This shows that \methodDA takes longer computation time with $\Delta G$ of larger size as analyzed in Section~\ref{subsec:deltaG}:
Larger size of $\Delta G$ leads to longer $L1$ length of the starting vector $\q_{\text{offset}}$ with longer computation time.
Likewise, longer $L1$ length of $\q_{\text{offset}}$ leads to higher $L1$ error as discussed in Section~\ref{subsec:deltaG}.
Still, the whole $L1$ norm errors are under the error bound $O(\epsilon/c)$ where $\epsilon$ is the error tolerance and $c$ is the restart probability.
%Note that we use the same error tolerance for different $\Delta G$ on each dataset, thus, $L1$ norm accuracies after modification with different sizes are similar.

\vspace{-2mm}
\subsubsection{Location of $\Delta G$}
\label{sec:exp-location}
To show the effects of location of $\Delta G$ on the performance of \methodDA, we estimate the running time varying the location of an edge to be deleted in a given graph.
We first calculate RWR scores among nodes and divide nodes evenly into $100$ groups in the order of RWR scores.
Then, we randomly sample $10$ nodes from each group.
For each sampled node $u$, we estimate the running time of \methodDA after deleting an edge $(u,v)$.
As shown in Figure~\ref{fig:delta:rwr}, when a modification happens around nodes with high RWR scores, \methodDA takes a long running time, but this rarely happens.
Intuitively, when a modification happens around high RWR score nodes, the higher offset scores are propagated, then the more steps would be needed to satisfy the given error tolerance.
This intuition is consistent with the theoretical result we showed in Section~\ref{subsec:deltaG}.
Sparse distribution around high RWR scores with long running time coincides with sporadic long running time observed by Ohsaka et al.~\cite{ohsaka2015efficient}.

\vspace{-7pt}
\section{Related Works}
\label{sec:related_works}
In this section, we review previous approaches to handle dynamic RWR problem.
Chien et al.~\cite{chien2004link} proposed the approximate aggregation/disaggregation method.
The method takes a small subset $S$ that contains the updated edge, and then renew RWR scores only in $S$.
Bahmani et al.~\cite{bahmani2010fast} applied the Monte-Carlo method~\cite{jeh2003scaling} on the dynamic RWR problem.
Their algorithm maintains R random-walk segments, and reconstructs any segments related to a graph modification.
%They analyzed time complexity in a random edge arrival order model.
Recently, score propagation models TrackingPPR and LazyForward were proposed by Ohsaka~\cite{ohsaka2015efficient} and Zhang~\cite{zhang2016approximate}, respectively:
{when a given graph is updated, scores that complement the changes are calculated at first, then propagated across the graph.}
Although sharing the same intuition, they differ in the initialization step and the propagation method.
While TrackingPPR propagates the scores immediately to all neighboring nodes, LazyForward modifies RWR values of the updated nodes at first then propagates the scores.
In \methodDA, our proposed method, calculating offset seed vector $\q_{\text{offset}}$ is at the initialization step.
The propagation methods used in the two models are Gauss-Southwell algorithm and Forward Local Push algorithm, respectively.
In each iteration, Gauss-Southwell algorithm and Forward Local Push algorithm propagate scores stored in a vertex which has the largest score,
while \methodDA propagates scores across whole vertices that could be reached in one hop.
%Their convergence conditions are different from 
TrackingPPR and LazyForward succeed in outperforming the previous approaches~\cite{chien2004link, bahmani2010fast} in both computation time and accuracy~\cite{ohsaka2015efficient, zhang2016approximate},
and show the effectiveness of propagation model on dynamic graphs.
However, none of them provide the guarantee of exactness 
{or rigid analysis of error bound. %and time complexity for general dynamic graphs.
Furthermore, Ohsaka et al. analyzed the running time only when edges are randomly and sequentially inserted, % for an edge-insertion model 
while Zhang et al. analyzed the running time only for undirected graphs.
Note that we provide exactness of \methodD, and time complexity and error bound for \methodDA in a generalized form.
\methodDA outperforms TrackingPPR and LazyForward in terms of speed and accuracy as shown in our experiments (Section~\ref{sec:exp-dynamic}).}
%This difference in propagating strategy leads to superior performance of \methodD than others as shown in our experiments (Section~\ref{sec:exp-dynamic}).

\vspace{-7pt}
\section{Conclusion}
\label{sec:conclusion}
We propose \methodD, a fast and accurate method for tracking RWR scores on a dynamic graph.
%\methodD is based on cumulative power iteration (\methodC) which interprets RWR problem as the propagation of scores from a seed node across a graph.
When the graph is updated, \methodD first calculates offset scores around the modified edges, propagates the offset scores across the updated graph, and then merges them with the current RWR scores to get updated RWR scores. 
We also propose \methodDA, a version of \methodD which regulates a trade-off between accuracy and computation time.
Among numerous dynamic RWR models based on score propagation, \methodD is the first model with rigid analysis of accuracy and running time in a generalized form.
Through intensive experiments, we show that \methodD and \methodDA outperform other state-of-the-art methods in terms of accuracy and computation time.
%Future works include extending \methodDA into a disk-based dynamic model to handle huge, disk-resident graphs. 

\bibliographystyle{ACM-Reference-Format}
\bibliography{BIB/myref}

\end{document}